%% file: splitALFpaper-main.tex
\documentclass[
onefignum,onetabnum]{siamart190516}

\usepackage{xspace}
\usepackage{dsfont}
\usepackage{longtable}
\newcommand*{\eg}{e.g.\@\xspace}
\newcommand*{\ie}{i.e.\@\xspace}
\newcommand{\etal}{{et al}.\@\xspace}

\input{splitALFpaper-shared}

\ifpdf
\hypersetup{
  pdftitle={Higher Order Auxiliary Field Quantum Monte Carlo Methods},
  pdfauthor={F. Goth}
}
\fi

\begin{document}

\maketitle

\begin{abstract}
The auxiliary field quantum Monte Carlo (AFQMC) method has been a workhorse in the field of strongly correlated electrons
for a long time and has found its most recent implementation
in the ALF package (\url{alf.physik.uni-wuerzburg.de}). The utilization of the Trotter decomposition
to decouple the interaction from the non-interacting Hamiltonian makes this method
inherently second order in terms of the imaginary time slice.
We show that due to the use of the Hubbard-Stratonovich transformation (HST)
a semigroup structure on the time evolution is imposed that necessitates
the introduction of a new family of complex-hermitian splitting methods for the purpose 
of reaching higher order.
We will give examples of these new methods and study their efficiency,
as well as perform comparisons with other established second and higher order methods
in the realm of the AFQMC method.
\end{abstract}



\section{Introduction}
The AFQMC method has a distinguished place among the numerical methods of solid state physics
for producing large scale benchmark data.
Its most recent implementation
is the Algorithms for Lattice Fermions (ALF) package (\url{alf.physik.uni-wuerzburg.de}).
The use of the Trotter product formula \cite{Trotter1959}
to decouple the interaction part $V$ of the Hamiltonian $H=T+V$ from the non-interacting kinetic energy $T$ makes this method
inherently second order in terms of the imaginary time discretization $\Delta \tau$.
This quality of approximation was initially sufficient and provides
a convergent approximation of all relevant quantities.

The wish for higher order approximations stems from studies of quantum criticality.
One application that has already lead to numerous Monte Carlo studies \cite{Assaad2013, Hesselmann2016, Huffman2017, ParisenToldin2015, Tada2020, Tang2018}
is research on quantum critical behaviour in 
fermionic Dirac systems which serve as models for the properties of graphene.
Of interest are the critical points of these systems where in the vicinity 
universal behaviour is expected that is not governed by the microscopic details and parameters
of the considered system but
determined by the fundamental symmetries of the Hamiltonian.
In these problems higher order methods could provide an advantageous
approximation and hence produce better critical exponents.
The AFQMC method has proven useful in these
studies since its favourable polynomial computational
complexity $\mathcal{O}\left( \beta N^3 \right)$, with particle number $N$
and inverse temperature $\beta$,
enables the study of large system sizes.
Another application is the study \cite{PhysRevB.98.235108} of quantum critical points of the non-linear sigma-model.
As observed by Wang \etal in \cite{Wang2020} the highly non-local interactions 
lead to hard to control discretization errors.

Hence it is of great interest to have alternatives to the established splitting methods that
enable a better control of the discretization error as well as an advantageous approximation of symmetries
and therefore motivates this search for higher order methods that are efficiently usable in the AFQMC method.

The AFQMC method dates back to a paper \cite{Blankenbecler1981} from Blankenbecler \etal
as a method for the simulation of lattice models subject to on-site interactions.
As it was soon found out, the algorithm is inherently numerically unstable 
due to the presence of long chains of multiplications of exponentiated matrices,
but publications \cite{BaiPropertiesofHubbardMatrices,Bai2011,Evertz2007} in the following years improved 
on this issue. Nowadays the algorithm has matured and is readily available in two
open source projects \cite{ALF2017, Quest2010} that allow the reliable simulation
of quantum lattice Hamiltonians as well as the extraction of relevant observables.
The AFQMC method can now be used for exploring the exponentially
growing many-particle configuration space in polynomial time while being 
numerically exact: investing more CPU time will decrease the error.

The possibility of operator splittings was shown with the Trotter product formula \cite{Trotter1959}
and work on difference schemes for partial differential equations lead to the symmetric second order splitting by Strang \cite{strang1968construction}.
Another very fertile ground for 
splitting methods has traditionally been the numerical solution of ordinary differential equations
most often from simulations of classical Hamiltonian systems as,
\eg molecular dynamics simulations or long-time astronomical calculations,
due to their superior conservation properties of very important physical quantities
like the total energy of the system. The prototypical example is the well-known leapfrog method - a variant of 
the Strang splitting - for classical Hamiltonian systems.
Higher order methods rose to prominence around the 1980s with
early forays beyond second order by Neri \cite{Neri1987} and Ruth \cite{Ruth1983}
and around the 1990s each field had their specific 
fourth order method: particle physics had the Forest-Ruth Integrator \cite{FOREST1990105},
astronomy had the higher order Integrators by Yoshida \cite{YOSHIDA1990262},
and solid state physics had the influential publication by Suzuki \cite{SUZUKI1990319}
that already explicitly mentions applications in Monte Carlo simulations of many body systems.
All publications note the connection and applicability to arbitrary Lie groups
and due to this the conservation of the symplectic structure of Hamilton's classical equations of motion was
established for these methods as well, a prerequisite for applications to Molecular Dynamics and Hybrid Monte Carlo
simulations \cite{Liu2004}.
For a detailed historical account as well as information on the further development of splitting methods 
we refer the reader to Ref.~\cite{mclachlan_quispel_2002}.
Although Suzuki has explicitly expressed the hope in \cite{SUZUKI1990319} that his methods
will be of use in Monte Carlo simulations,
in the realm of solid-state physics and the AFQMC method in particular 
there has to our knowledge not been a widespread use of higher order approximations
to the partition function $Z$ or other physical quantities,
except for occasional uses of the Strang splitting \cite{He2019, Assaad2013}.
Now with the recent impetus in solid-state physics due to the aforementioned
challenges observed in fermionic quantum criticality we aim to explore the issues of 
higher order methods.

The intended audience of this paper consists of numerical mathematicians as well as physicists
alike and hence the paper will begin in
\cref{sec:derivation} with a pedagogical derivation of the AFQMC method in the language of differential equations.
In detail we start in \cref{subsec:splittingmethods}
with a short introduction into splitting methods geared towards our purposes.
\Cref{subsec:hub} outlines the physical setting and gives a very brief definition of the Hubbard model.
Afterwards, we apply this in \cref{subsec:bss}
to derive a splitting method dependent
representation of the partition function.
With that we can move on to define the actual Monte Carlo dynamics in \cref{subsec:mc} but since this is not 
the focus of this paper we emphasize the results and point the reader towards the relevant literature.
The derivation will close with a discussion of the systematic error in \cref{subsec:error}.

Having the mathematical preliminaries, \cref{sec:comp} will focus on numerically
comparing the performance of various splitting methods.
We first compare second order methods that have found success in 
the Hybrid Monte Carlo method in \cref{subsec:secondorder} and we give numerical evidence
that their superior numerical behaviour can also be observed in the AFQMC method.
After that the content on higher order methods begins with \cref{subsec:classical}
where we numerically study certain classical fourth order methods.
Due to the Goldman-Kaper theorem \cite{Goldman1996, BLANES200523, Chin2004} we are faced with a bad 
sign problem and are forced into the complex plane in \cref{subsec:cmplx}.
\Cref{subsec:symrealv} considers complex symmetric methods with one positive set of coefficients.
\Cref{subsec:hermrealv} focuses on hermitian methods and first
analyzes the structure of their order conditions.
Then we introduce the new family of hermitian splitting methods
where one set of coefficients is entirely real. We give numerical 
evidence that, so far, this class provides a very good average sign of a simulation
while still being of high discretization order.
We conclude with an outlook 
about various directions for future research in \cref{sec:outlook}
and finish with the conclusion in \cref{sec:conclusions}.

In the appendix we collect various results that are peripheral to the subject of the main text.
In \cref{sec:ghquad} we have a derivation of the discrete HST
that easily scales to higher orders and to our knowledge has not been given in the literature before in that form.
\Cref{app:euler} gives a detailed error analysis of the error of observables in the AFQMC method,
and \cref{app:hocbm} a small numerical case study 
where we apply higher order methods to the checkerboard decomposition to efficiently evaluate matrix exponentials.
We conclude with a small numerical investigation in \cref{app:suppHST} that explores how so far unused 
low order approximations from \cref{sec:ghquad}
can be utilized in the AFQMC method.

The coefficients of the new class of hermitian methods can be found in \cref{app:table}.
See \cref{app:hocbm} for the notation and
where to find big overview tables in different representations online.

All simulations carried out in this paper have been done with a fork
of the ALF package version 1.0 \cite{ALF2017} called splitALF
\footnote{splitALF can be obtained from \url{https://git.physik.uni-wuerzburg.de/fgoth/splitALF}.}.

\section{Derivation}
\label{sec:derivation}
\subsection{Splitting methods}
\label{subsec:splittingmethods}
Given a vector $\psi$ from a Hilbert space $\mathcal{H}$ and the autonomous differential equation
\begin{equation}
 \dot{\psi}(\tau) = - H \psi(\tau)
 \label{eq:ode}
\end{equation}
with the Hamilton operator $H$ we can write the solution of \eqref{eq:ode} as
\begin{equation}
 \psi(\tau) = U_H(\tau, 0) \psi(0)
\end{equation}
with the time evolution operator
\begin{equation}
 U_H(\tau, 0) = \exp(-\tau H).
\end{equation}
In all that follows we will suppress the dependence of the time evolution $U(\tau, 0)$ on the initial time and just write $U(\tau)$.
If now $H$ has an additive substructure, where we will restrict ourselves for the purpose of this paper to the case 
of two non-commuting constituents,
\begin{equation}
 H = T + V,
\end{equation}
with a kinetic energy part $T$ and an interaction part $V$, then for small enough $\tau$
a first order approximation is given by the expression
\begin{equation}
\begin{split}
 U_H(\tau) = e^{-\tau (T+V)} &= \exp(-\tau T)\exp(-\tau V) + \mathcal{O}(\tau^2) \\
  & =  U_T(\tau)U_V(\tau) + \mathcal{O}(\tau^2)
 \end{split}
\end{equation}
where it is now hoped that the individual flows of $T$ and $V$, $U_T$ and $U_V$ are in a to be precised way 
easier to calculate than the full problem $U_H(\tau)$.
Of course we  do not have to stop at a single application of $T$ and $V$ but we can study splittings
that consist of multiple of these stages, and we obtain the
splitting method $X_s p$ defined by the tuple of parameters \met{X}{s}{p} $=(\vec{t}, \vec{v})$
\begin{equation}
 U^{X_s p}_H(\tau) = \prod_{k=1}^{s+1} U_T(t_k \tau) U_V(v_k \tau) + \mathcal{O}(\tau^{p+1})
 \label{eq:splitting}
\end{equation}
where $s+1$ counts the number of stages, \ie the number of evaluations of $T$ and $V$.
$p$ defines the order of the method in the sense that for a particular
method \met{X}{s}{p} and a given vector norm $\lVert \bullet \rVert$ the following holds
\begin{equation}
 \lVert \psi(t) - \psi^{\text{X}_sp}(t) \rVert = \lVert U_H(\tau) \psi(0) - U^{\text{X}_sp}_H(\tau) \psi(0) \rVert = \mathcal{O}(\tau^{p+1}).
\end{equation}
The symbol X will be chosen to denote a property of the evolution and
we will just write $\text{X} p$ where only the order, or even only X in contexts where the precise splitting is 
not relevant.

For the method X to be at least first order accurate, the consistency equations,
\begin{equation}
 \sum \limits_k^{s+1} t_k = \sum \limits_k^{s+1} v_k = 1,
\end{equation}
have to hold. The remaining degrees of freedom of a particular method can be used for various purposes,
the most common one being an increase of the order by enforcing a cancellation of higher order terms \cite{YOSHIDA1990262, SUZUKI1990319, McLachlan1995, Thalhammer2008}.
Other goals are a minimization of the abstract truncation error at a particular order \cite{McLachlan1995, BLANES2002313},
in the context of Hybrid Monte Carlo a minimization of the energy error \cite{BLANES2014, 2019arXiv191203253C},
an increase of the stability interval with the help of the linear stability \cite{Blanes2008}
or the mitigation of errors due to approximations in the flows $U_T$ or $U_V$ \cite{BLANES201358}.

To apply the notation we consider the most well-known method that utilizes two stages but has one application of $U_V$,
which is the Strang, or symmetric, splitting given by
\met{S}{1}{2} $= ((1/2,1/2), (1,0))$. Hence
\begin{equation}
 U^{\text{S}_1 2}_H(\tau) = \exp\left(-\frac{\tau}{2}T\right) \exp(-\tau V)  \exp\left(-\frac{\tau}{2}T\right) + \mathcal{O}(\tau^3).
\end{equation}
Here $s=2$ and the method is of order $p=2$, a result that can be easily obtained from the property that the local truncation error 
is an odd function of $\tau$ for a time reversal symmetric method \cite{YOSHIDA1990262}.

We will call a method time reversal symmetric and denote it by the symbol S if 
\begin{equation}
 U^\text{S}_H(-\tau) U^\text{S}_H(\tau) = 1.
\end{equation}
This is equivalent to $v_{s+1}=0$ and a palindromic set of coefficients with $t_{s-k+2} = t_k$ and $v_{s-k+1} = v_k$.
We will call a method hermitian or self-adjoint and denote it by the symbol {CH} if 
\begin{equation}
 \left[ U^{\text{CH}}_H(\tau) \right]^\dagger = U^{\text{CH}}_H(\tau).
\label{eq:hermitianevolution}
\end{equation}
This is equivalent to $v_{s+1}=0$ and a hermitian set of coefficients with $t^*_{s-k+2} = t_k$ and $v^*_{s-k+1} = v_k$.
We require this distinction since, although both notions coincide if a method $X$ is chosen from the domain of real numbers,
it does not coincide in the complex domain.
Except for the basic Euler method all methods considered in this paper are at least hermitian and hence
$v_{s+1}=0$.
Now $s$ counts the true number of applications of $U_V$, which enables us to use the \met{X}{s}{p} notation in a manner that is in line with the literature.

Since in general a method is not generically stable for all values of $\tau$ we have to concatenate
a succession of $L_\tau$ time steps of size $\Delta \tau$,
\begin{equation}
 U_H(\tau) = \prod \limits_{n=1}^{N\tau} U_H(\Delta\tau).
\end{equation}
Inserting the method dependent flows $U^\text{X}_H$ we deduce for the global error accumulated for the time evolution to the point
$\tau$ in time steps $\Delta \tau$
\begin{equation}
\begin{split}
 U_H(\tau) & = \prod \limits_{n=1}^{N\tau} U^\text{X}_H(\Delta \tau) + \mathcal{O}(\Delta \tau^p). 
 \end{split}
 \label{eq:approxfullflow}
\end{equation}
We emphasize that this notion of stability with respect to the chosen step size is different to the
numerical instability that is commonly associated with the AFQMC method \cite{BaiPropertiesofHubbardMatrices,Bai2011,Evertz2007}.

\subsection{The Hubbard model}
\label{subsec:hub}
The physical setting is determined by the Hamilton Operator $H$ that is acting on the fermionic Hilbert space
and is in our case given by the Hubbard model,
a model often studied using the AFQMC method (see \cite{ALF2017,Varney2009} and references therein.)
The Hubbard Hamiltonian is a very simple model that considers electrons on a lattice with a
Hamiltonian consisting of two parts,
\begin{equation}
\label{eq:hubtwoops}
 H=T+V.
\end{equation}
The kinetic energy part $T$ models electrons hopping to their nearest neighbours.
The potential energy $V$ is an on-site interaction with coupling strength $U$ between electrons of different spin.
From this we have the representation which is used in ALF,
\begin{align}
H & = T - U \sum \limits_{i=1}^{L_N} V_i^2 \label{eq:Hub}\\
  & = \sum \limits_{i,j=1}^{L_N} \sum \limits_\sigma c_{i\sigma}^\dagger T_{ij} c^{\phantom{\dagger}}_{j\sigma} - \frac{U}{2} \sum \limits_{i=1}^{L_N} \left(\sum \limits_\sigma \sigma n_{i\sigma}\right)^2
\end{align}
with $U>0$, $[V_i, V_j] = 0$, $n_{i\sigma} = c^\dagger_{i\sigma} c^{\phantom{\dagger}}_{i\sigma}$, and
$c^{\phantom{\dagger}}_{i\sigma}$ ($c^{\dagger}_{i\sigma}$)
denoting the annihilation (creation)
operators that annihilate (create) an electron in state $i$ with spin $\sigma$.
$\sigma$ enumerates the two spin directions
$\uparrow, \downarrow$.
We study this model on a two-dimensional regular square lattice of linear length $L_N$ with periodic boundary conditions.
This fixes the hopping matrix $T$ to $T_{ij} = \delta_{i,j+1}+\delta_{i+1,j}$.

\subsection{The AFQMC method}
\label{subsec:bss}
For our purposes the partition function $Z$ of a physical system determined by the Hamilton operator $H$ is defined 
as the trace taken in the many particle Fock space over the fermionic degrees of freedom of the time evolution operator $U_H$,
\begin{equation}
 Z  = \Tr\left( U_H(\beta)\right) = \Tr\left(e^{-\beta H}\right)
\end{equation}
with inverse temperature $\beta$.
We split up the propagation
into $L_\tau$ time slices
\begin{equation}
 Z= \Tr(U_H(\beta)) = \Tr \left( (U_H(\Delta \tau))^{L_{\tau}} \right).
\end{equation}
Due to eq.~\cref{eq:hubtwoops} we know that we only have two operators and hence
we can approximate the exact $U_H$ with the approximate flow of a method \met{X}{s}{p} from \eqref{eq:splitting}
to obtain
\begin{equation}
\label{eq:split}
\begin{split}
Z &= \Tr \left( \prod_{\tau=1}^{L_{\tau}} \prod_{k=1}^{s+1} U_T(t_k \Delta \tau) U_V(v_k \Delta \tau) \right) + \mathcal{O}(\Delta \tau^p)\\
  &= \Tr \left( \prod_{\tau=1}^{L_{\tau}} \prod_{k=1}^{s+1} e^{ - \Delta \tau t_k T} e^{  \Delta \tau v_k U \sum \limits_{i=1}^{L_N} V_i^2} \right) + \mathcal{O}(\Delta \tau^p)\\
  &= Z^{\text{X}_sp} + \mathcal{O}(\Delta \tau^p)
 \end{split}
\end{equation}
with a suitably chosen method such that the order $p$ is achieved.
We take a short break to appreciate that we now have a splitting method dependent approximation $Z^{\text{X}_sp}$ 
to the value of the true partition function $Z$.
Here we will apply the AFQMC method \cite{Blankenbecler1981} to formulate a Monte Carlo method but it is also conceivable to use
the Hybrid Monte Carlo method \cite{PhysRevB.36.8632, Beyl2018}
or Langevin dynamics \cite{Duane1985} on the approximation $Z^\text{X}$.
To reduce the interaction which is quartic in the number of fermion operators to a quadratic form we use
on every lattice site $i$, every imaginary time slice $\tau$ and for every stage $s$
a discrete approximation to the HST of order $q$ (for a basic derivation see \cref{sec:ghquad}),
and obtain
\begin{equation}
 \exp(v_k \Delta \tau U V_i^2) = \sum_{l_{i,\tau,k}=1}^{N(q)} w({l_{i,\tau,k}}) e^{\sqrt{v_k \Delta \tau U } x({l_{i,\tau,k}}) V_i} + \mathcal{O}\left(\Delta \tau^q\right).
 \label{eq:discHST}
\end{equation}
$w({l_{i,\tau,k}})$ denotes the weights and $x({l_{i,\tau,k}})$ the abscissas
of a suitable chosen quadrature rule of order $q=p+1$ and hence $N(q)$ integration nodes.
On every timeslice and lattice site we obtain an expression that only involves terms that are quadratic in the fermionic operators
and we arrive for the partition function at
\begin{equation}
 Z = \sum_{C\in\mathcal{C}} \left[ \prod_{i,\tau,k=1} w(l_{i,\tau,k}) \right] \Tr \left( \prod_{\tau}^{L_\tau} \prod_{k=1}^{s+1}  e^{ - \Delta \tau t_k T} \prod_{i=1}^{L_N} e^{ \sqrt{\Delta \tau v_k U} x(l_{i,\tau,k})  V_i} \right) + \mathcal{O}\left(\Delta \tau^p\right).
\end{equation}
$\sum \limits_{C\in\mathcal{C}}$
denotes a summation over the entire space $\mathcal{C}$ of configurations $C$ that we will later on sample stochastically.
$\mathcal{C}$ is finite dimensional and discrete:
\begin{equation}
 \mathcal{C}=\{ l_{i,\tau,k} \}\quad:\quad  1 \leq i \leq L_N, 1 \leq \tau \leq L_\tau, 1 \leq k \leq s+1
\end{equation}
and each integer $l_{i,\tau,k}$ takes on values determined by the order $q$ of the  Gauss-Hermite quadrature.
This configuration space of auxiliary fields is for computational purposes and not directly related 
to any physically interpretable quantity as \eg in the canonical example of the Ising-Model \cite{newman1999monte}.
Note that the time evolution of the interaction operator is now determined by $\sqrt{v_k \Delta \tau U}$.
For a well-defined Monte Carlo sampling over positive probability densities
we want the quantities under the square root to be real and hence the interaction does not form a group anymore but only a semigroup.
Using the identity proven in \cite{Evertz2007} valid for matrices $A^{(k)}$ with entries $A^{(k)}_{ij}$
\begin{equation}
 \Tr\left(\prod_k \exp\left(\sum \limits_{ij} c^\dagger_i A^{(k)}_{ij} c^{\phantom{\dagger}}_j \right) \right)= \det\left(1 + \prod_k e^{A^{(k)}} \right)
\end{equation}
that enables us to map the trace over the fermionic Fock space on the left hand side
to a determinant of a product of exponentiated square, complex matrices on the right hand side,
we finally arrive at
\begin{align}
Z & = \sum\limits_{C\in \mathcal{C}} P(C) + \mathcal{O}\left(\Delta \tau^p\right) \text{ with } \label{eq:partfuncfinal} \\
P(C) & =   \left[ \prod_{i,\tau,k} w(l_{i,\tau,k}) \right] \det \left( 1 + \prod_{\tau=1}^{L_\tau} B^{X_s}_{\tau} \right) \text{ and } \\
B^{\text{X}_s}_\tau & = \prod_{k=1}^{s+1}  e^{ - \Delta \tau t_k T} \prod_{i=1}^{L_N} e^{ \sqrt{ \Delta \tau v_k U} x(l_{i,\tau,k})  V_i}.
\end{align}
We have introduced for later the probability density $P(C)$.
Its positive-definiteness for different physical systems has to be checked on a case-by-case 
basis.
This concludes the derivation of an expression for the partition function that is amenable to stochastic evaluation.
\subsection{Metropolis Monte Carlo and observables}
\label{subsec:mc}
We will very briefly mention in this section the relevant steps for the Metropolis update scheme and follow closely Ref.~\cite{Evertz2007}.
We construct a Markov chain Monte Carlo method to generate samples to evaluate the sum in \eqref{eq:partfuncfinal}.
Starting from a configuration $C$ (initially randomly chosen integration nodes), a move consists of
changing the value of one integration node $x_{i,\tau,k} = x_n$ to another $x_{i,\tau,k} = x_n'$ to obtain a new 
configuration $C'$. We accept the new configuration according to the Metropolis-Hastings rule
\begin{equation}
 r_{MH} = \min\left(\frac{P(C')}{P(C)}, 1\right).
 \label{eq:mh}
\end{equation}
This gives an ergodic Markov chain Monte Carlo method that enables us to generate configurations on which to
measure observables according to
\begin{equation}
\label{eq:obs}
 \langle O \rangle = \frac{\Tr\left( U_H(\beta,0) O \right)}{\Tr\left( U_H(\beta,0) \right)} = \sum \limits_C P(C) O(C)
\end{equation}
where $O(C)$ is the numerical value of an observable $O$ evaluated on the configuration $C$.
Since each configuration $C$ is equivalent to a non-interacting system, Wick's theorem can be used to 
find expressions for arbitrary physical observables. We will not get into details here, but instead the interested
reader can find expressions for observables as well as methods for the efficient calculation of the ratio
of determinants occuring in eq.~\eqref{eq:mh}
in the specialized literature on the AFQMC method \cite{BaiPropertiesofHubbardMatrices, ALF2017, Evertz2007}.
Nevertheless, for later reference in this paper we have to mention the sign, or more generally, phase problem
which occurs if in \eqref{eq:obs} $P(C) > 0$ is violated and is hence negative, or even an arbitrarily complex number.
To mitigate this issue and at least have the possibility of generating a valid Markov chain one employs
a reweighting procedure. 
This requires 
the notion of the sign $\sigma$ of a configuration
\begin{equation}
 \sigma(C) = \frac{P(C)}{|P(C)|}.
\end{equation}
With that we obtain for an observable \cite{ALF2017} the modified equation
\begin{equation}
\label{eq:obssign}
 \langle O \rangle =\frac{\sum \limits_C |P(C)| \sigma(C) O(C)}{\sum \limits_C |P(C)| \sigma(C)}
\end{equation}
with the now real and non-negative probability distribution $|P(C)|$. The denominator in that expression
\begin{equation}
 \sigma = \sum \limits_C |P(C)| \sigma(C)
\end{equation}
gives rise to the notion of an average sign of a particular simulation. If $P(C) > 0 ~\forall~ C$ obviously means $ \sigma = 1$ and hence the quotient
in \cref{eq:obssign} is well-behaved, but if the sign deteriorates the statistical error bars will be blown up by forming the quotient.

\subsection{Error propagation}
\label{subsec:error}
With \eqref{eq:partfuncfinal} we have a representation of the partition function that is of order $p$, but will this carry over to the
observables?
The value of an observable $O$ is given by the expression
\begin{equation}
 \langle O \rangle = \frac{\Tr \left( U_H(\beta) O\right)}{Z}.
\end{equation}
Inserting the representation of the exact time evolution $U_H$ in terms of the approximate time evolution 
from \eqref{eq:approxfullflow} in the form
\begin{equation}
 U_H(\beta) = U^{\text{X}}_H(\beta) + \Delta \tau^p D
\end{equation}
with a defect $D= \sum \limits_{i=0}D_i \Delta \tau^i, D_0\neq 0$ we obtain
\begin{equation}
 \langle O \rangle = \frac{ \Tr \left( U^\text{X}_H(\beta) O\right) + \Delta \tau^p \Tr(D O)  }{Z^\text{X} + \Delta \tau^p \Tr(D)  }.
\end{equation}
Assuming that $|\Delta \tau^p \Tr(D)|$ is small we can use $(1+x^p)^{-1} \approx 1-x^p$ and obtain
\begin{align}
 \langle O \rangle &=\frac{ \Tr \left( U^\text{X}_H(\beta) O\right)}{Z^\text{X}} + \Delta\tau^p
 \left(\frac{\Tr(D O)}{Z^\text{X}} - \frac{\Tr(D)}{Z^\text{X}}  \frac{\Tr(U^\text{X}_H(\beta) O)}{Z^\text{X}}\right) + \mathcal{O}\left(\Delta \tau^{2p}\right)\nonumber \\
 &= \langle O \rangle^{\text{X}} + \mathcal{O}\left( \Delta \tau^p\right) \label{eq:errobs}
\end{align}
with the method dependent value of the observable
\begin{equation}
\label{eq:methoddependenterror}
 \langle O \rangle^{X} = \frac{ \Tr \left( U^\text{X}_H(\beta) O\right)}{Z^\text{X}}.
\end{equation}
This shows that the observable $O$ inherits its $\Delta \tau$ scaling behaviour to leading order from the partition function.
A more detailed analysis for the error propagation in the particular case of the Euler splitting as used in ALF-1.0 is given in \cref{app:euler}.

\section{Comparison of second and higher order methods in the AFQMC algorithm}
\label{sec:comp}
In this section we will study the scaling behaviour with respect to the step size $\Delta \tau$
of an important physical quantity,
the total energy
\begin{equation}
E^X(\Delta \tau) =\langle H \rangle^X (\Delta \tau)
\end{equation}
on a very small test case where analytical results are available.
According to eq. \cref{eq:errobs} we see that the systematic deviation with respect to $\Delta \tau$
enables us to access the leading order error term. Depending on the precision that we are able to achieve
for a particular class of simulations, a double logarithmic plot will expose the power law behaviour.
Where appropriate, we will give measurements of the average sign \cref{eq:obssign}.
In the case of simulations in complex arithmetic we will suppress any occuring artificial imaginary parts of 
the energy and the average sign, since they are small for the considered parameters.
\Cref{subsec:hermrealv} has some more details for that case.
The intrinsic method dependent parameters of ALF that control its numerical stability
have been chosen such that numerically stable simulations were performed with all methods.

\subsection{Second order methods}
\label{subsec:secondorder}
In addition to the well-known Strang splitting \met{S}{1}{2} \cite{strang1968construction}, we have the minimum-norm Integrator \met{S}{2}{2}
by McLachlan \cite{McLachlan1995}. It 
was the first integrator that utilized the additional degrees of freedom provided by the additional third stage
for a minimization of the leading order error term and was thereby able to improve on the Strang splitting.
Later on, the method was put forward for the integration of Hamilton's equation of motion in \cite{Omelyan2002,Takaishi2006}.
Additionally, we have the family of second order methods given by Blanes \etal in \cite{BLANES2014}.
They are designed to minimize the energy error in HMC simulations and thereby optimize the acceptance rate.
The associated methods \met{SE}{n}{2} are symmetric, of second order and have $n=2,3,4$ evaluations of $U_V$.

For the second order simulations we have used a fourth order approximation of the HST
in \eqref{eq:discHST}.

\begin{figure}
\label{fig:compsecondorder}
 \centering
 \includegraphics[width=\linewidth]{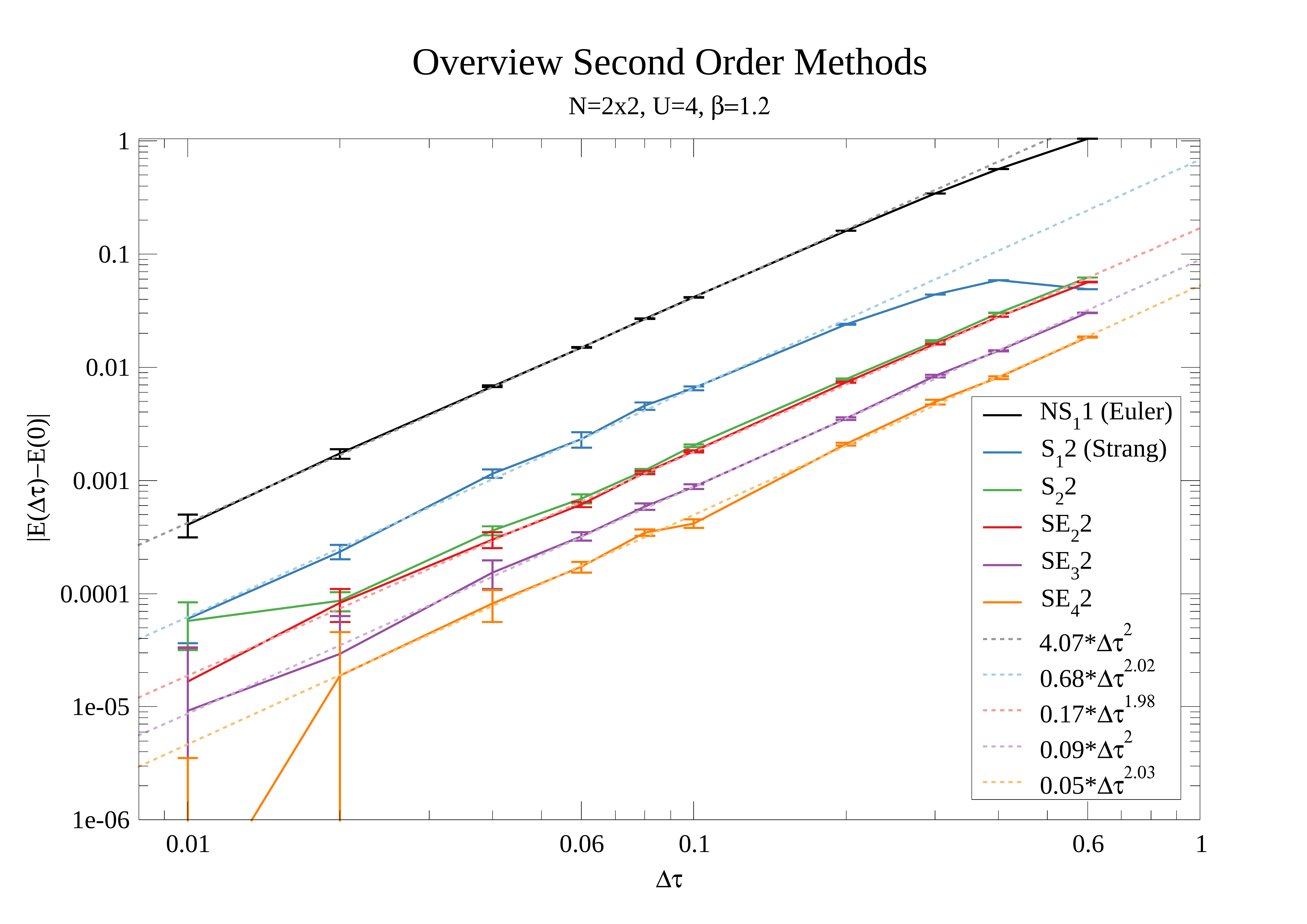}
 \caption{Scaling behaviour with respect to the step size $\Delta \tau$ of the considered symmetric second order methods
 studied on the systematic deviations of the total energy, $E(0) - E(\Delta \tau)$. The exact value $E(0) = -0.89200074$
 is known from Exact Diagonalization calculations.
 }
\end{figure}
\Cref{fig:compsecondorder} gives an overview of the scaling of the second order methods with respect to the step
size $\Delta \tau$. Since all involved splitting coefficients are positive the average sign is one.
We see that the non-symmetric Euler approximation \met{NS}{1}{1} is of second order but with a big prefactor in 
comparison to the symmetric Strang splitting \met{S}{1}{2},
an observation that already occured in \cite{Assaad2013}.
This behaviour and the analytic reason is studied in more detail in \cref{app:euler}.
The improved methods \met{SE}{n}{2} provide substantial benefits 
over the conventional Strang splitting. Combined with the fact that they can be easily expanded to 
splittings consisting of an arbitrary number of operators (similar to what is done in \cref{app:hocbm})
we highly recommend to use e.g. at least the \met{SE}{2}{2} method for the construction of the approximate 
partition function. Whether methods involving more stages provide a tangible benefit depends 
on the computing time costs associated with a stage.

\subsection{Classical higher order methods}
\label{subsec:classical}
Having studied alternatives to the trusty Euler and Strang splittings in the realm of second order methods
we now strive to increase the order of the used splittings.
A classical result about the structure of integration methods known as the Goldman-Kaper bound
tells us that any method beyond order two only utilizing the flows $U_T$ or $U_V$
involves negative coefficients \cite{Sheng1989,SUZUKI1990319, Chin2004}. We reproduce the theorem
here as given in Ref.~\cite{BLANES200523}:
\begin{theorem}[Goldman \cite{BLANES200523, Goldman1996}]
 If $p$ is a positive integer such that $p \geq 3$, then, for every $p$th-order method \met{X}{s}{p} with $s$ any finite
 positive integer, one has
 \begin{equation}
  \min \limits_{1\leq i \leq s+1} t_i < 0 \quad \text{and} \quad \min_{1\leq j \leq s+1} v_j < 0.
 \end{equation}
\end{theorem}
The fact that negative coefficients are unavoidable will pose us some challenges since the flow
$U_V$ does not form a full group that can be arbitrarily moved backwards in time.
This will yield imaginary contributions in the Monte Carlo sampling and hence induce an
artificial sign problem although the Hubbard model would be sign problem free for the considered parameters.
One way to circumvent this would be the technique of modified potentials \cite{Takahashi1984, casas2008splitting}
where higher order commutators are included into the time evolution \eqref{eq:splitting}.
But recent results have shown that this technique cannot be extended beyond order four \cite{Auzinger2019}
and the required commutators can generally not be cast back into the form 
used in ALF.\\
To illustrate the problem we will first study some classical examples with real coefficients that belong into this category and, instead of focusing on the scaling
behaviour with respect to $\Delta \tau$, we will focus on the average sign.
The methods we study are a selection of classical ones and currently recommended ones.
We start the classical selection with \met{S}{3}{4}, the original Forest-Ruth algorithm \cite{FOREST1990105} first derived in a technical 
report by Neri \cite{Neri1987} and put into a larger framework by Yoshida \cite{YOSHIDA1990262}.
It has the minimum number of three stages that are required to have enough degrees of freedom to satisfy the order conditions
for a symmetric fourth order method.
Then we have a representative from a family of Suzuki \cite {SUZUKI1990319}, \met{SS}{5}{4},
which has been studied in much detail by McLachlan \cite{McLachlan2002}, and is constructed as a symmetric composition of suitably chosen symmetric methods.
An early instance of a minimum error fourth order method was determined by McLachlan \cite{McLachlan1995}
by a minimization of the associated truncation error and this resulted in the method \met{S}{5}{4}.
We close with two methods, a symmetric fourth order method \met{S}{6}{4} and, just for comparison,
a symmetric sixth order method, \met{S}{10}{6}, that have both been given by Blanes \etal in Ref.~\cite{BLANES2002313}.
They have also been derived by a minimization of the first non-vanishing error term and are shown to produce 
smaller truncation errors than all preceding fourth order methods.
For the fourth order method we use a sixth order approximation and for the sixth order method we use an eigth order
approximation to the HST in \eqref{eq:discHST}.
From this we observe a slight decrease in the acceptance ratio. See \cref{app:suppHST} for a more detailed study of the acceptance rate.
\begin{figure}
\label{fig:compclassicalfourthorder}
 \centering
\includegraphics[width=0.48\linewidth]{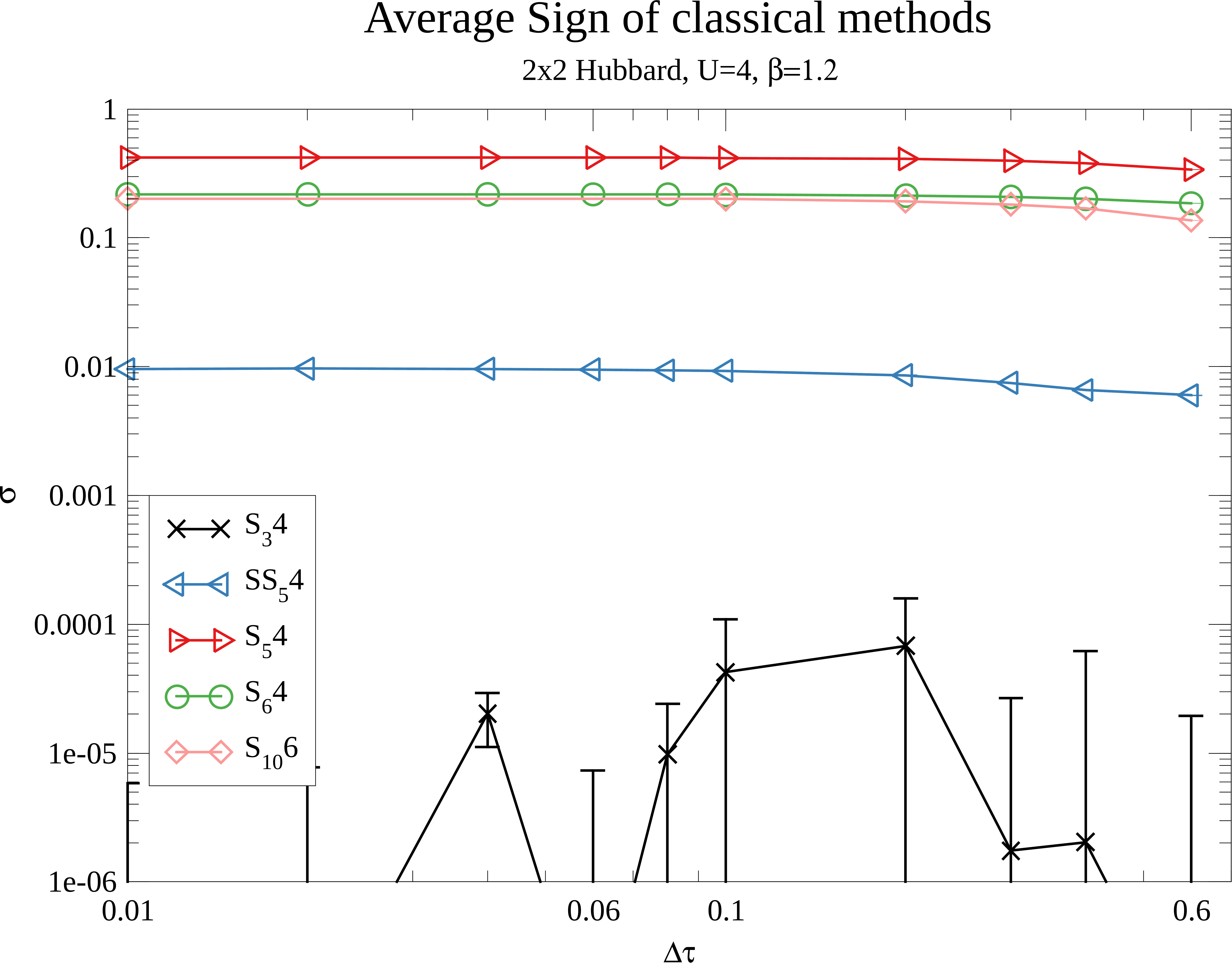}
\hfill
 \includegraphics[width=0.48\linewidth]{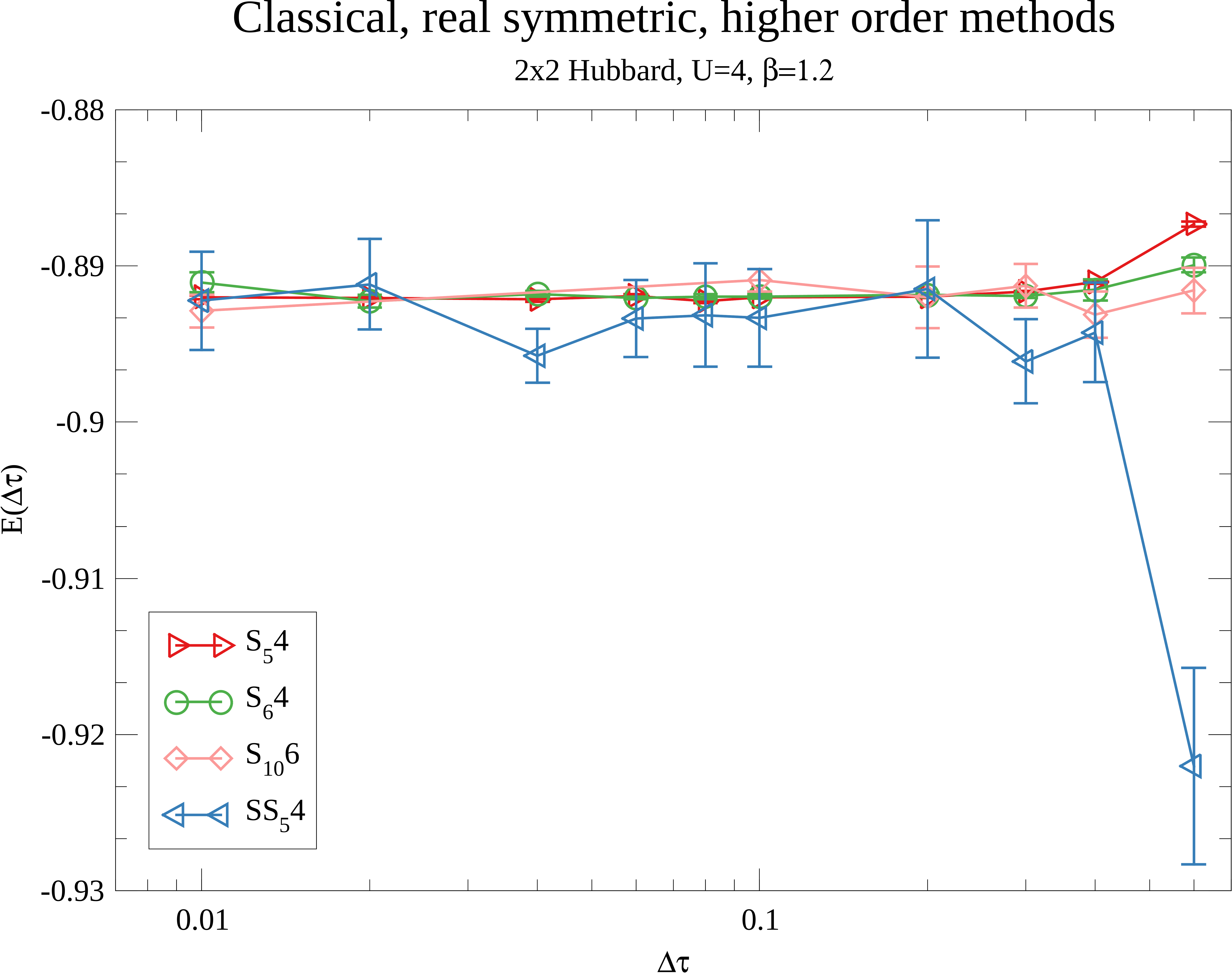}
 \caption{
 Comparison of some classical fourth order methods with respect to the approximation of the energy and
 the resulting  average sign.
 The left panel shows the average sign of a simulation and we observe that all methods produce an appreciable sign problem.
 The right panel shows us that all of them are still able to reproduce
 the correct value of the energy within their error bars.
 Method \met{S}{3}{4} is left out from the right plot since it gives
 such a tremendously bad average sign ($\sigma < 10^{-5}$) as can be seen in the left.}
\end{figure}

We attribute the abysmal performance of all considered methods so far to the fact that 
the negative coefficients of $\vec{v}$ have to propagate through the HST \cref{eq:discHST}
which leads to complex phase factors turning up in the Monte Carlo sampling.
If we focus on the fourth order methods we note that all of them have a single negative coefficient in the coefficient
set $\vec{v}$ and the average sign shown in \cref{fig:compclassicalfourthorder} seems to be directly 
related to the magnitude of it. \met{S}{5}{4} with $v_2=-0.1$ gives an average sign almost twice as good as
\met{S}{6}{4} which has $v_2=-0.14$. The bottom end is marked by \met{S}{3}{4} which has a diminishing sign and has $v_2= -1.7$.
\footnote{See \cref{app:hocbm} on how to find tables with the coefficients.}
Nevertheless, the appearance of an appreciable sign problem for such a tiny test-case does not provide solid
ground for the usual simulations of large systems.

\subsection{Complex splittings}
\label{subsec:cmplx}
In the previous subsection we have seen that the semigroup structure imposed by the HST for 
$U_V$ leads to serious numerical difficulties in the form of a sign problem which stems from the
negative coefficients contained in the set of coefficients $\vec{v}$.
The possibility of complex solutions to the order conditions has already been noted by Suzuki \cite{SUZUKI1990319}
and he has determined the hermitian fourth order method, \met{CH}{4}{4}. A number of families that include this method
have been given in Ref.~\cite{Hansen2009}.
In the complex plane we now have the issue that a hermitian method will not be time reversal symmetric and vice versa.
Refs.~\cite{blanes2010splitting, Hansen2009, SUZUKI1990319} have given the well-known hermitian third order method \met{CH}{2}{3}, and in \cite{Hansen2009} we find the time reversal symmetric, non-hermitian \met{CS}{3}{4} method of order four.
In a generic setting some care has to be taken since the analytic continuation of the time evolution operator $U(\tau)$
to the full complex plane, $U(z)$ need not be well-defined for all values of $z$ but instead there
could be a restriction to a sector $\Sigma_\phi = \{z \in \mathbb{C}: \arg(z) < \phi \}$.
Hansen \etal \cite{Hansen2009} have shown that under some additional technical assumptions a method 
\met{X}{s}{p}
with complex coefficients $(\vec{t}, \vec{v})$ in the sector $\Sigma_\phi$
will retain their classical order $p$, derived via formal Taylor series expansions,
even in the context of unbounded operators.

\begin{figure}
\label{fig:complexmethods}
 \centering
 \includegraphics[width=0.48\linewidth]{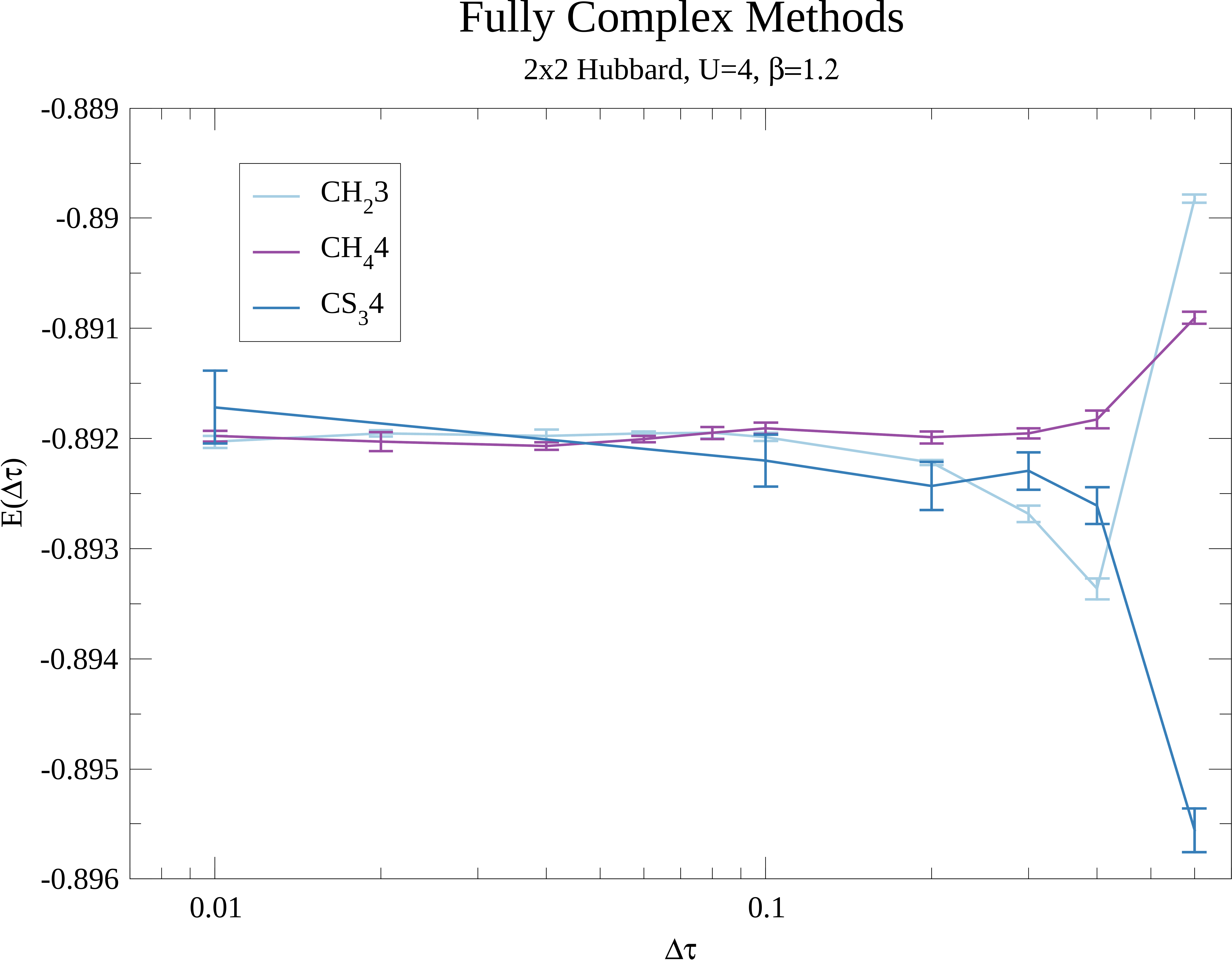}
 \hfill
 \includegraphics[width=0.47\linewidth]{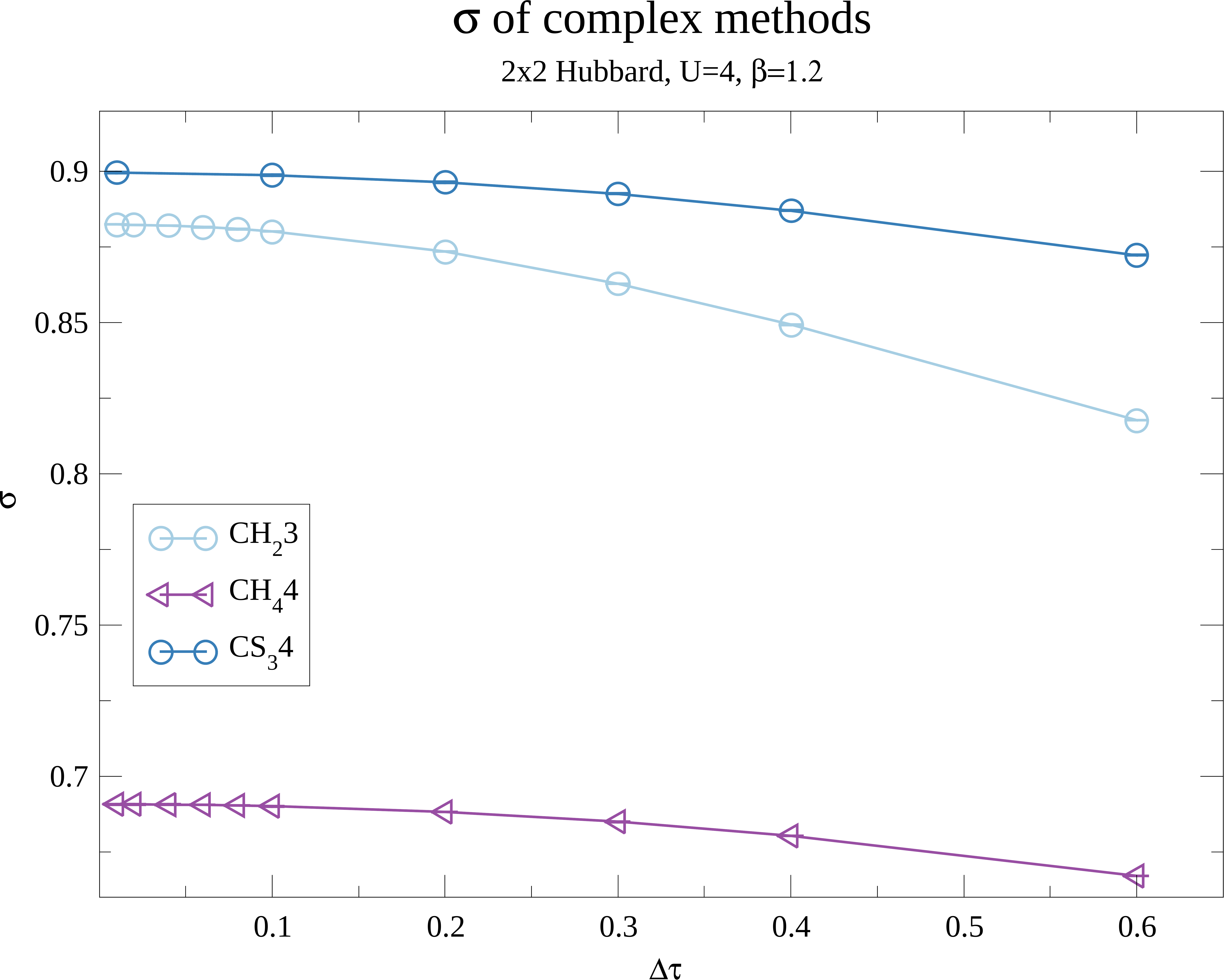}
 \caption{
 Comparison of the energy behaviour and the average sign of methods that involve fully complex splitting coefficients.
 }
\end{figure}

\subsection{Complex splittings with positive $v_i$}
\label{subsec:symrealv}
Having seen that complex methods can provide results with an average sign that is superior to the real-valued methods,
although we pass complex coefficients $\vec{v}$ through the HST, it is natural to ask
 whether methods exist that have real, positive $\vec{v}$ and hence might have an improved average sign.
An early positive indication to their existence has been given by Castella \etal in Ref.~\cite{castella2009splitting}
and optimized symmetric methods from this class have been published by Blanes \etal \cite{Blanes2012}.
In this section we study the fourth order method \met{CSR}{4}{4} and sixth order method \met{CSR}{16}{6}
from \cite{Blanes2012}. We note that these methods are time reversal symmetric but not hermitian.
\begin{figure}
\label{fig:complexrealvmethods}
 \centering
 \includegraphics[width=0.47\linewidth]{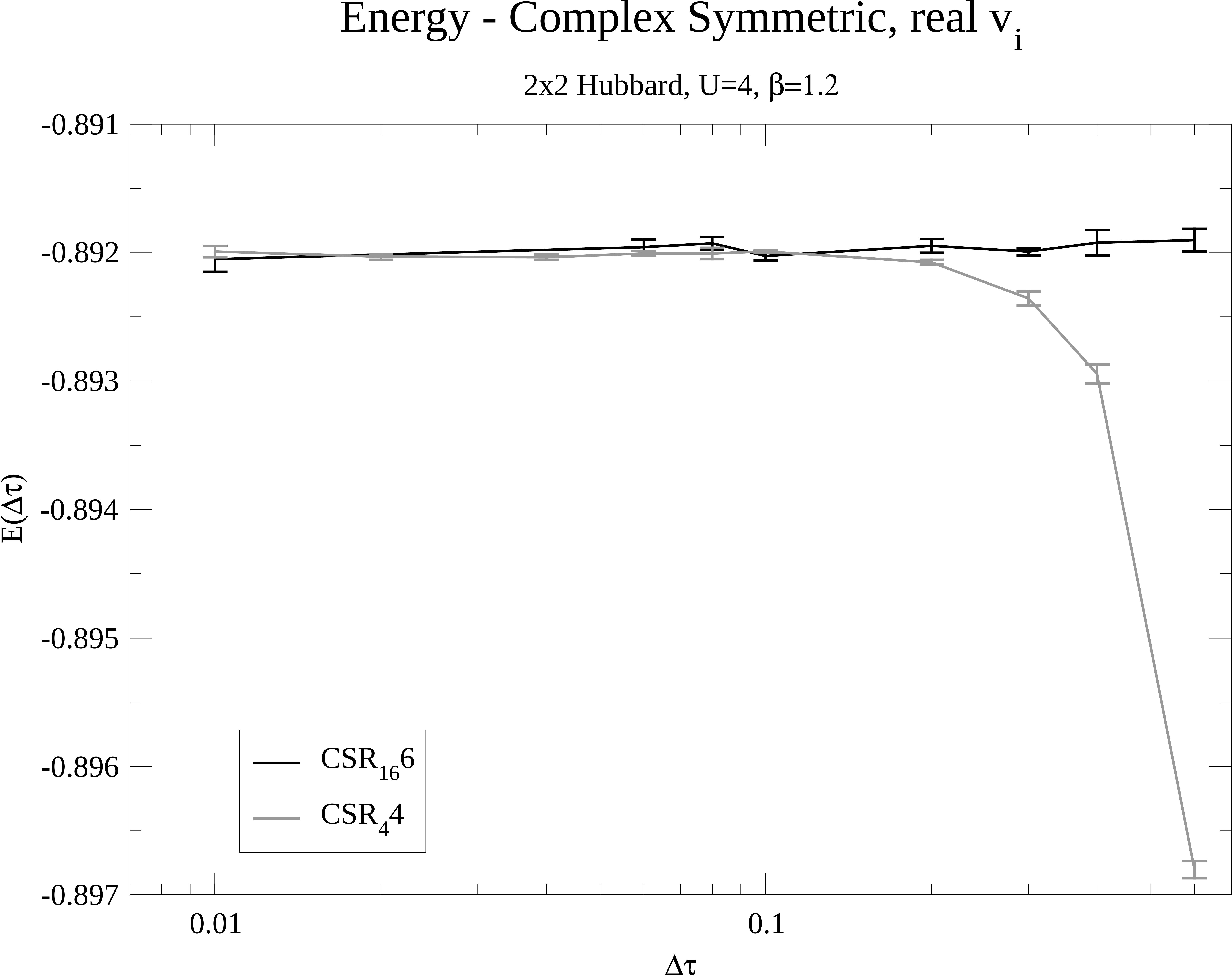}
 \hfill
 \includegraphics[width=0.48\linewidth]{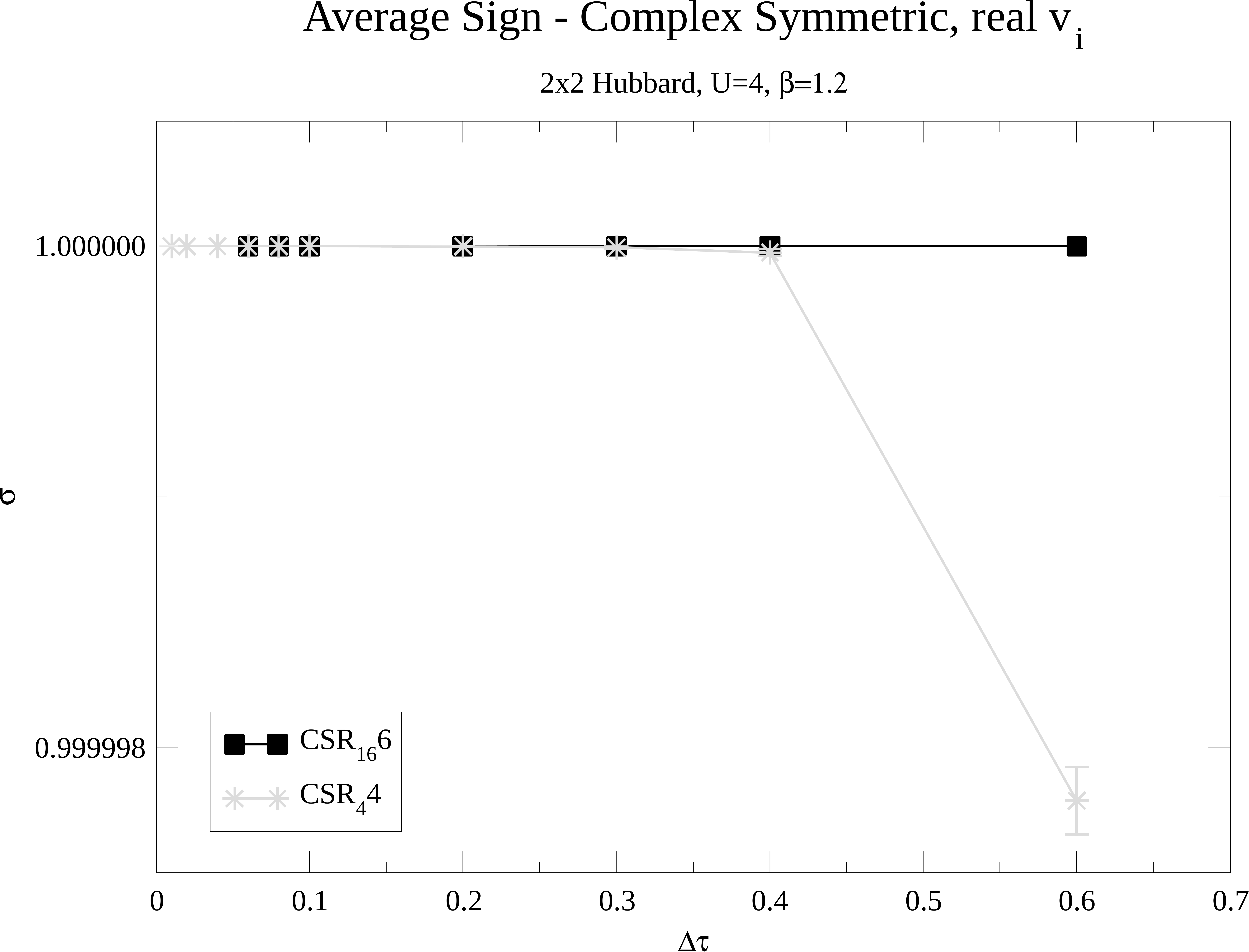}
 \caption{
 Comparison of symmetric, but non-hermitian methods where $\vec{v}$ is real and positive.
 The left panel shows convergence to the expected value of the energy and the right panel shows
 that the average sign is almost indistinguishable from $\sigma=1$.
 }
\end{figure}
From \cref{fig:complexrealvmethods} we see that we are able to accurately reproduce the expected value of the energy
with an average sign that is almost indistinguishable from unity.

\subsection{Hermitian splittings with positive $v_i$}
\label{subsec:hermrealv}
Given the success of methods where $v_i > 0$ a serious drawback remains:
the resulting decomposition is time reversal symmetric but not hermitian.
To properly expose this issue in measurable quantities we have to study slightly more complicated,
$\tau$ dependent observables.
A particular relevant observable to see this problem is the local Green's function $G(\tau)$.
Its definition is
\begin{equation}
 G(\tau) = \sum \limits_{i,j,\sigma,\sigma'} \langle c^\dagger_{i\sigma}(\tau) c_{j\sigma'} \rangle.
\end{equation}
Using the definition of the time evolution of an operator $O(\tau) = U_H(\tau) O U_H(-\tau)$
and the property that the Hamiltonian $H$ is self-adjoint and hence real diagonalizable, $H = \sum_n E_n | n \rangle \langle n |$,
with the real eigenvalues $E_n$ and eigenstates $| n \rangle$, we find for $G(\tau)$ the representation
\begin{equation}
 G(\tau) = \frac{1}{Z} \sum \limits_{n,m} e^{-\beta E_n} e^{\tau(E_n - E_m)} |\langle n | \sum \limits_{i\sigma} c_{i\sigma}| m\rangle |^2
\end{equation}
and hence $G(\tau) \in \mathbb{R}$.
If the approximate time evolution operator $U^{X_s p}_H(\tau)$ from eq.~\cref{eq:splitting}
breaks hermiticity
$G(\tau)$ will have unphysical imaginary parts.
\begin{figure}
\label{fig:greenscomp}
 \centering
 \includegraphics[width=\linewidth]{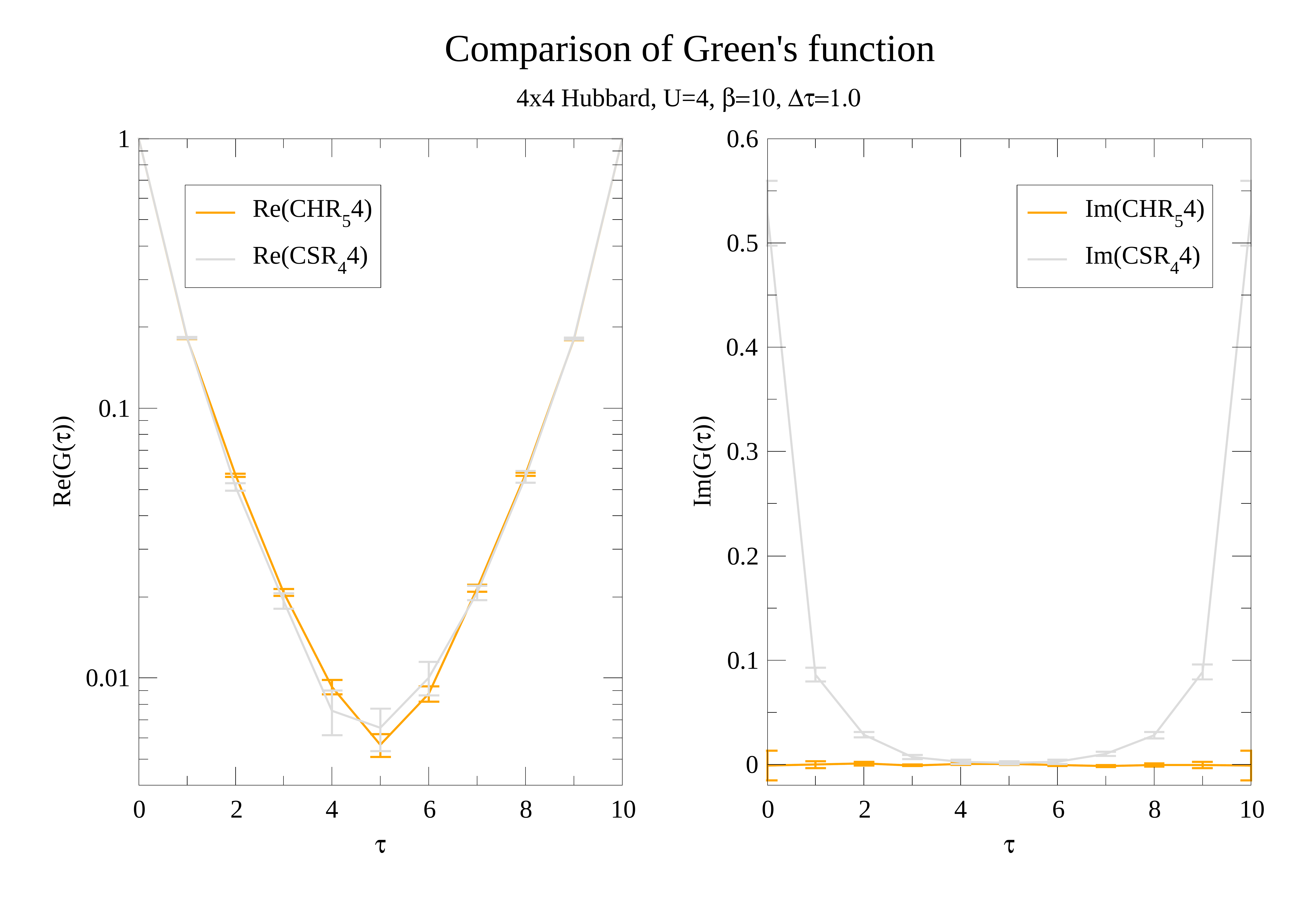}
 \caption{This is the local Green's function $G(\tau)$ for the Hubbard model. The left side depicts the real part and the 
 right side depicts the imaginary part. While the real part of the two methods looks almost identical,
 the non-hermitian \met{CSR}{4}{4} method produces pretty big unphysical imaginary parts whereas the hermitian 
 method \met{CHR}{5}{4} does not and, instead, is vanishing within its error bars.}
\end{figure}
\Cref{fig:greenscomp} compares the complex symmetric method \met{CSR}{4}{4} with the hermitian method
\met{CHR}{5}{4}.
on  $G(\tau)$, and we see exactly this happening.
We observe that while the real part of $G(\tau)$ turns out very similar for both methods,
the imaginary parts differ wildly. Only the hermitian method produces a vanishing imaginary part of
$G(\tau)$ as expected.
Hence it is clear that we need to find splitting methods that preserve
the hermiticity while additionally having a positive real coefficient set $\vec{v}$ similar as in the previous section.
On top of that, Ref.~\cite{blanes2010splitting} has shown that hermitian methods with complex coefficients
exhibit superior stability properties compared to time reversal symmetric ones: if applied to 
the harmonic oscillator they do not give exponentially diverging solutions.
To our knowledge no specimen of this particular type of method has been published in the literature so far.
To fill the gap we used the equations
\footnote{Jupyter and Mathematica notebooks to derive the equations are available at \url{https://captainsifff.github.io/pySplitting/}.}
from \cite{Thalhammer2008} to derive our own methods subject to the additional conditions
$v_i>0$ $\forall i$ and $t_i = t_{s-i+2}^*$.
\subsubsection{Counting order conditions}
In order to successfully solve the order conditions it is of prime importance to 
obtain the minimal set of independent equations.
Ref.~\cite{Thalhammer2008} gives an expression for the order conditions,
but the equations obtained by this approach are not all independent due to the presence of the Jacobi identity in a Lie-algebra.
As outlined in \cite{BLANES201358} a minimal set of order conditions that are in contrast generically 
independent can be found by considering a subset of these equations defined by the so-called Lyndon words over the alphabet 
\{T,V\}.
As detailed by an argument in \cite{BLANES201358} the set of equations obtained this way is isomorphic
to the classical approach that relies on the Baker-Campbell-Hausdorff (BCH) formula \cite{hall2015lie} to
obtain the order conditions.
With the help of the Lyndon words we are now in a position to directly get the equations
whose number is equal to the known number $n_{\text{NS}}$ (\cref{tab:nrorderconds}) of independent order conditions \cite{casas2008splitting}
that have to be considered at each order $p$.

\begin{table}
\centering
\begin{tabular}{|c|cccccc|}
\hline
 $p$ & 1 & 2 & 3 & 4 & 5 & 6 \\
 \hline
 $n_{\text{NS}}$ &2 & 1 & 2 & 3 & 6 & 9 \\
 \hline
\end{tabular}
\caption{From Ref.~\cite{casas2008splitting} we have the number $n_{\text{NS}}$ of order conditions at each order $p$ for a non-symmetric method.
We have two conditions for $p=1$ since we also count $\sum_i t_i = 1$. For a symmetric method the conditions for even orders
are automatically fulfilled.
}
\label{tab:nrorderconds}
\end{table}

Ordinarily, if we take the order conditions into the complex domain, we would expect a doubling
of the number of equations since real and imaginary part contribute one equation each,
but further simplifications can be obtained for hermitian methods with two self-adjoint operators $T$ and $V$.
Applying the BCH formula repeatedly to \cref{eq:splitting} one obtains the following representation of a method $CH p$ of order $p$,
\begin{equation}
 U^{\text{CH}}_H(\tau) = \exp\left(\sum \limits_{i=1}^{p} \tau^i \sum \limits_{j=1}  P^{\text{CH}}_{i,j}(\vec{t}, \vec{v}) C_{i,j}  + \mathcal{O}\left( \tau^{p+1} \right)\right),
\end{equation}
where $C_{i,j}$ denotes repeated commutators of $T$ and $V$ at order $i$,
and the index $j$ labels the occuring commutators, \ie
the basis element,  at a particular order.
$P^{CH}_{i,j}(\vec{t}, \vec{v})$ are the method dependent polynomials.
Imposing the requirement of hermiticity, eq.~\cref{eq:hermitianevolution}, and comparing the arguments of 
the exponentials at each order $\tau^i$ we find that
\begin{equation}
 \sum_j P^{CH}_{i,j}(\vec{t}, \vec{v}) C_{i,j} = \sum_j \left(P^{CH}_{i,j}(\vec{t}, \vec{v})\right)^* C^\dagger_{i,j}
\end{equation}
has to hold.
For two self-adjoint operators we have $C^\dagger_{i,j} = (-1)^i C_{i,j}$ and hence the requirement
\begin{equation}
 P^{CH}_{i,j}(\vec{t}, \vec{v}) -(-1)^i \left(P^{CH}_{i,j}(\vec{t}, \vec{v})\right)^* = 0
\end{equation}
for the polynomials follows.
By that argument we see that the polynomials for the order conditions are either purely imaginary 
or purely real and hence the anticipated doubling of the degrees of freedom is canceled.
This structure helps in directly using the requirements $v_i > 0$ and $t_i = t_{s-i}^*$,
so that order conditions for $\Re(t_i), \Im(t_i)$ and $v_i$ in the real domain can be obtained.
Hence we see that a complex hermitian method has the same number of real degrees of freedom as a real, non-symmetric method.

\subsubsection{Particular solutions}
In the following we report on some particular solutions that we considered. No attempt at an optimization in any particular
kind has been done.
From \cref{tab:nrorderconds} we see that a third order method has to fulfill five order conditions, hence 
it can be satisfied by a system with three real stages. Among the possible solutions we picked
the method \met{CHR}{3}{3}.
For a fourth order method we need to fulfill eight conditions and hence require five stages. This yields
a one-parameter family from which we picked method \met{CHR}{5}{4}.
At fifth order nine stages are required to fulfill 14 order conditions.
From the resulting one-parameter family we chose the method \met{CHR}{9}{5}.

\begin{figure}
\label{fig:scalingwithexternalparameters}
 \centering
 \includegraphics[width=0.48\linewidth]{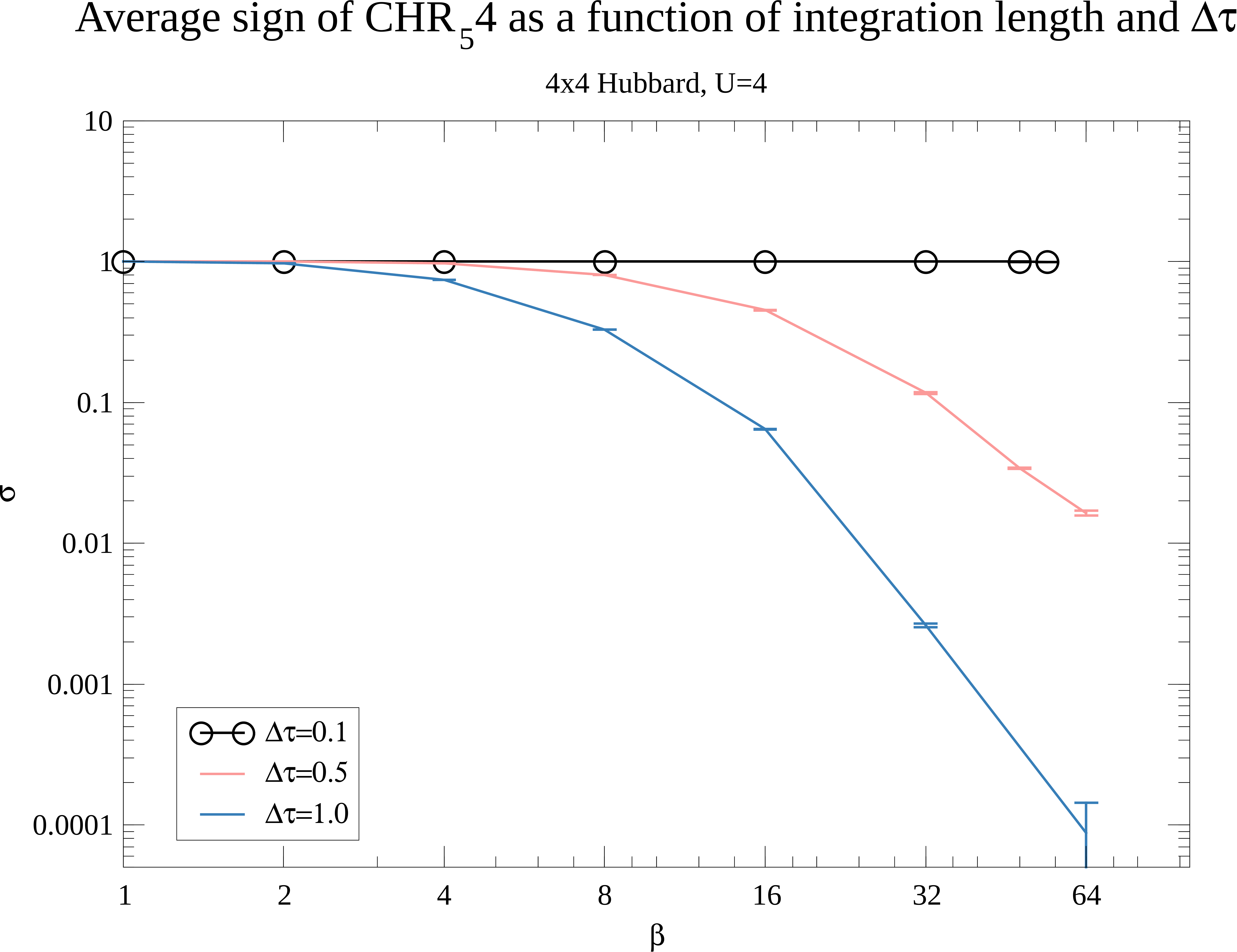}
 \hfill
 \includegraphics[width=0.49\linewidth]{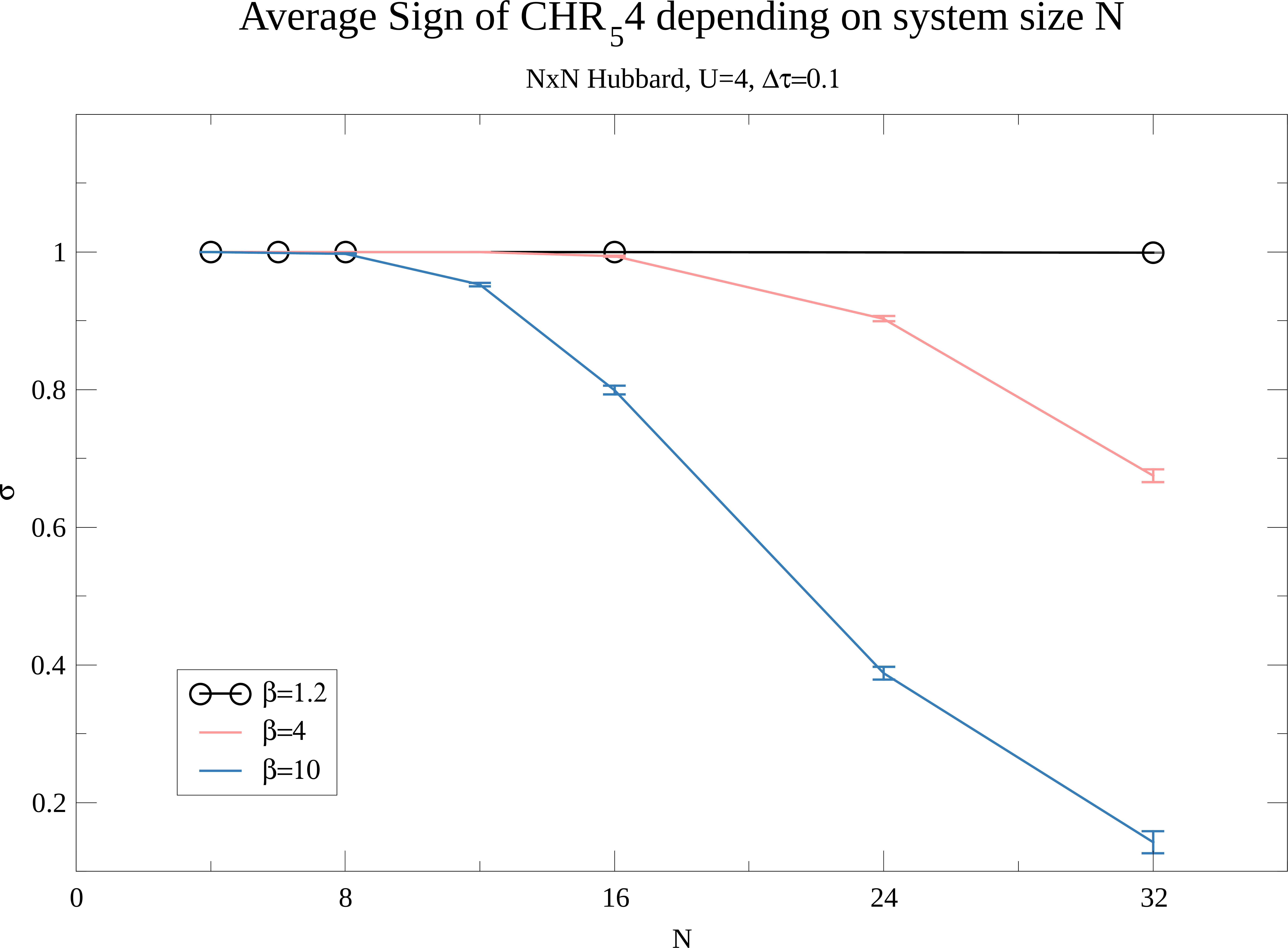}
 \caption{
 In the left panel we have the decay of the average sign for a $4\times4$ system at $U=4$ with respect to $\beta$ for three 
 different discretizations.
 We see a strong dependence of the average sign on $\Delta\tau$ and obtain an exponential decay with respect to $\beta$ for large $\Delta\tau$.
 The right panel shows the decay of the average sign with respect to the linear system size $N$ for three values of $\beta$.
 We observe again a behaviour reminiscient of an exponential decay.
 }
 \end{figure}

The final sixth order method $\text{CHR}^+_{15} 6$ that we considered has 15 stages to fulfill 23 order conditions.
We note that all given hermitian methods also fulfill the property
$\Re(t_i)>0$ and hence could be suitable for the study of evolution equations \cite{blanes2010splitting, Hansen2009}.
\met{CHR}{3}{3} and \met{CHR}{5}{4} will be the two methods that we study deeper in this section;
methods \met{CHR}{9}{5} and \met{CHR}{15}{6} will be considered briefly for comparison afterwards.
\begin{figure}
\label{fig:hermitianrealvmethods}
 \centering
 \includegraphics[width=0.48\linewidth]{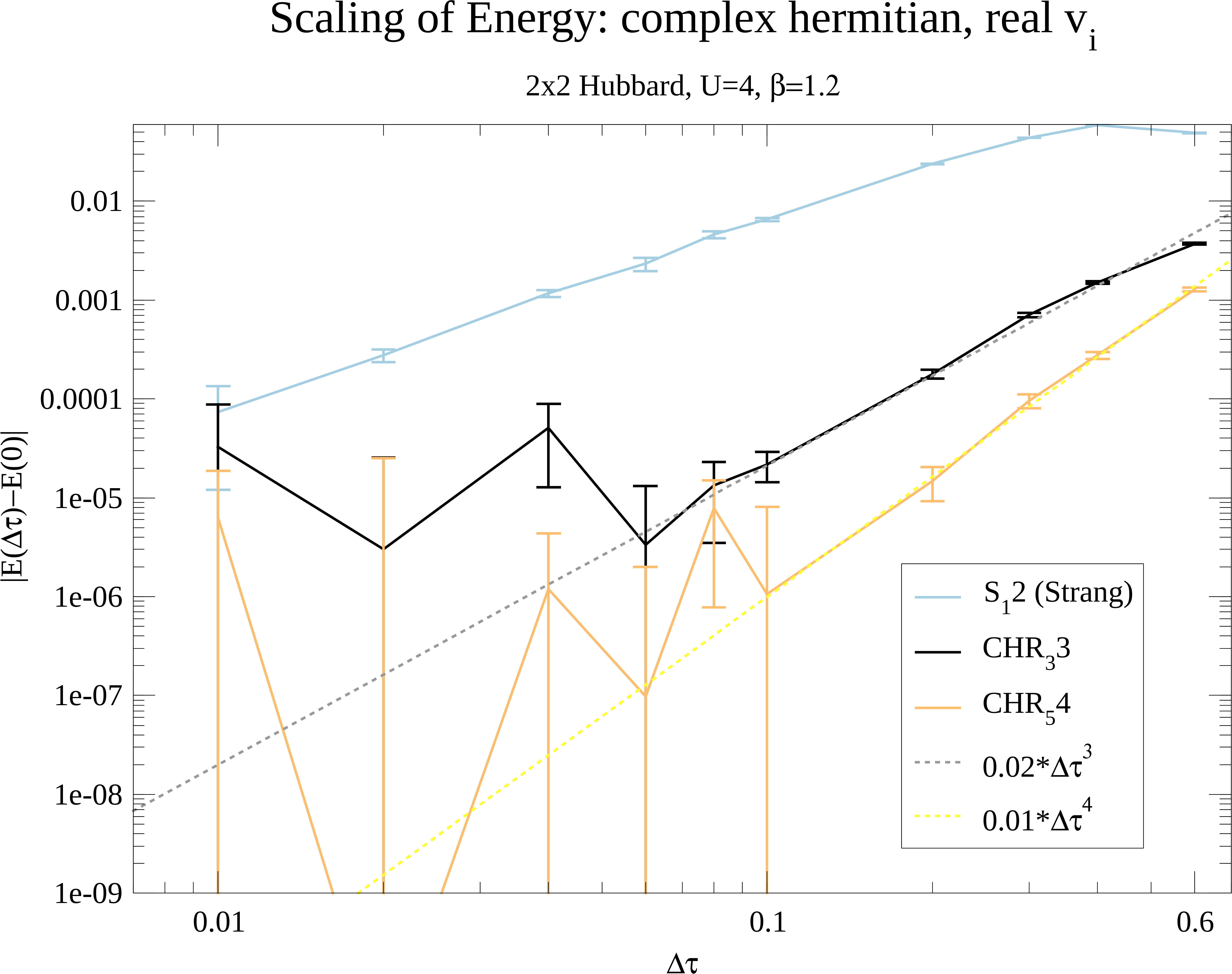}
 \hfill
 \includegraphics[width=0.48\linewidth]{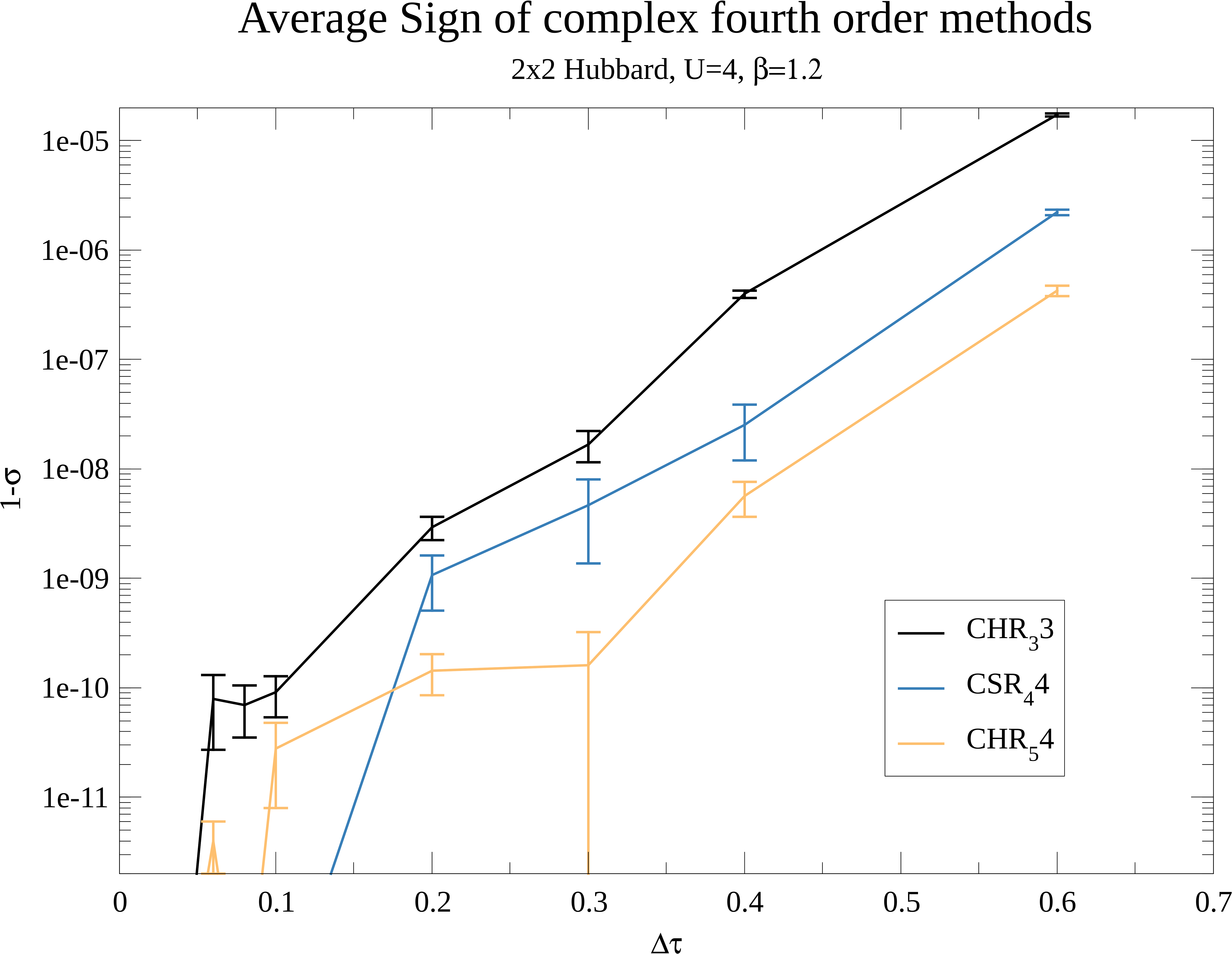}
 \caption{
 Comparison of the hermitian, but not time reversal symmetric methods \met{CHR}{3}{3} and \met{CHR}{5}{4}
 where $\vec{v}$ is real and positive. We see that within their error bars the methods approach the expected polynomial
 decay.
 We note that the resolution of the systematic $\Delta \tau$ error poses a numerical 
challenge for the high-order methods. This is another reason why this section only considers third and fourth order methods.
 Note that in the right panel we give $1-\sigma$ instead of $\sigma$ and we see that smaller $\Delta \tau$ leads to an 
 improved average sign.
 }
\end{figure}
\Cref{fig:hermitianrealvmethods} gives the scaling behaviour of the error as well as the behaviour of the average sign.
We see that the proposed methods are able to reach their expected scaling behaviour
and introduce a slight sign problem on the toy example.
Hence it is of interest to study the behaviour for larger $\beta$ and bigger system sizes as done in \cref{fig:scalingwithexternalparameters}.
\Cref{fig:scalingwithexternalparameters} studies the average sign with respect to a variation of the step-size $\Delta \tau$,
the inverse temperature $\beta$ and the system size $N$. We see that with an increase of $\beta$ and $N$ the average
sign worsens exponentially.
We observe that increases in the step size lead to a worse average sign, a property that has been noted for the AFQMC
algorithm in \cite{Mikelsons2009}.

\subsection{Dependence of average sign on the order of the splitting}
Higher integration orders usually confer a better approximation.
And better approximations are often associated with an improved average sign.
To shed light on this we set up an experiment with six different complex higher order methods that we considered 
in this paper and chose $\Delta\tau$ appropriately so as to renormalize their computational effort.
For this experiment we set the number of nodes in the HST to eight and considered the $4\times4$ cluster at $U=4$ and
$\beta=18$. The stabilization parameter $N_\text{Wrap}$ of ALF was set such that $N_\text{Wrap}*\Delta \tau = 0.9$ which leads
to stable simulations for all methods.
\begin{figure}
\label{fig:signvsorder}
 \includegraphics[width=0.8\linewidth]{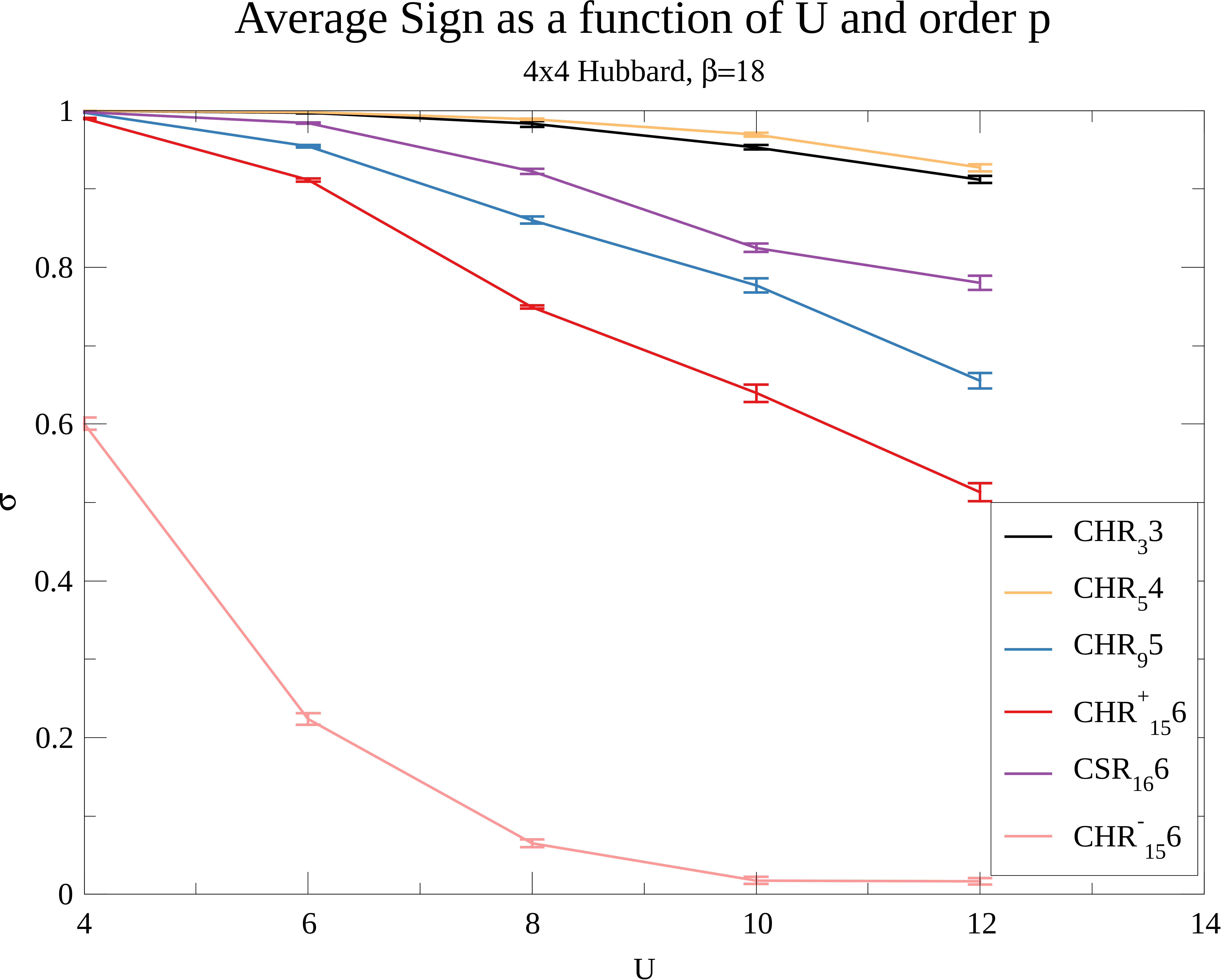}
 \caption{We compare complex higher order methods with respect to $U$ and their order $p$. We see that the average sign
 $\sigma$ decays for all methods with increasing $U$.
 What stands out is the behaviour of the method \met{CHR$^-$}{15}{6} which already 
 starts at $U=4$ with an average sign only half as big as the other methods.
 }
\end{figure}

As can be seen from \cref{fig:signvsorder}, the average sign only slightly worsens with increased order $p$.
The ranking among the methods follows roughly an increase in Hansen \etal's method angle \cite{Hansen2009}:
\begin{definition}[Method Angle \cite{Hansen2009}]\label{def:ma} The method angle $\phi_X$ for a method $X_sp = (\vec{t}, \vec{v})$ is defined as
\begin{equation}
 \phi_X = \max \left(|\arg(\vec{t})|, |\arg(\vec{v})|\right),
\end{equation}
with $\arg(\vec{y}) = \{\arg(y_i)\}_{i=1}^s$ and $\arg(y_i) = \arctan\left(\Im(y_i)/\Re(y_i)\right)$.
\end{definition}
\begin{table}
\centering
\begin{tabular}{|c|cccccc|}
\hline
 $X$ & \met{CHR}{3}{3} & \met{CHR}{5}{4} & \met{CHR}{9}{5} & \met{CHR$^+$}{15}{6} & \met{CSR}{16}{6} & \met{CHR$^-$}{15}{6} \\
 \hline
 $\phi_X$ &$0.167\pi$ & $0.195\pi$ & $0.419\pi$ & $0.405\pi$ & $0.343\pi$ & $0.895\pi$ \\
 \hline
\end{tabular}
\caption{Overview of method angles for the considered methods.
}
\label{tab:metangles}
\end{table}
From \cref{tab:metangles} we see that methods with good average sign seem to have a smaller method angle than methods with bigger method angle.
Additionally, the difference between \met{CHR$^+$}{15}{6} and \met{CHR$^-$}{15}{6} begs for an explanation.
We believe that this is due to the fact that \met{CHR$^+$}{15}{6} as well as all other considered \met{CHR}{s}{p} methods of lower order additionally have positive real parts for $\vec{t}$ whereas \met{CHR$^-$}{15}{6} has not.

\section{Outlook}
\label{sec:outlook}
\subsection{Beyond the Hubbard model}
So far we have studied the Hubbard Model which has been considered in this paper as having two 
non-commuting quantities, $T$ and $V$. 
We have seen that the semigroup structure imposed on $U_V$ due to the HST leads to issues for higher order
methods that can be alleviated with the help of the special hermitian higher order methods of \cref{subsec:hermrealv}.
Of course it is of interest what we can do if we have more than two operators,
as $H=\sum_i H_i$. First we will reorder and relabel them as 
\begin{equation}
 H=\sum \limits_i^{N_T} T_i  + \sum \limits_j^{N_V} V_j = T + V
\end{equation}
where now every $T_i$ generates a group and every $V_j$ is an interaction that requires decomposition with the HST.
The $T_i$ could \eg stem from a checkerboard decomposition for which we show an example in \cref{app:hocbm}.
Then we first split these two groups of operators with a hermitian
fourth order method, \eg \met{CHR}{5}{4}, and have 
\begin{equation}
 U_H(\tau) = \prod_{i_{TV}} U_T(t^{(CHR4)}_{i_{TV}} \tau) U_V(v^{(CHR4)}_{i_{TV}} \tau).
\end{equation}
Hence the effect of the non-commutativity of $T$ and $V$ has been suppressed until fourth order.
Now we need to split $U_T$ and can choose any suitable high-order method for $N_T$ operators(see \cref{app:hocbm})
with an order strictly bigger than four, \eg a symmetric sixth order method.
This gives
\begin{equation}
 U_T(t^{(CHR4)}_{i_{TV}} \tau) = \prod \limits_{i_T} \prod_{k=1}^{N_T} U_{T_k}(c^{(S6)}_{i_T} t^{(CHR4)}_{i_{TV}} \tau)\prod_{k=N_T}^1 U_{T_k}(d^{(S6)}_{i_T} t^{(CHR4)}_{i_{TV}} \tau)
\end{equation}
where we have used a symmetric of symmetric decomposition with coefficients $c_i$ and $d_i$ similar as outlined in \cref{app:hocbm}.
There are plenty of methods to choose from since $U_T$ is a full group.
Now the effect of the non-commutativity among the $T_i$ has been supressed until sixth order.
A similar construction is to be employed for the $V_j$, but we are limited to second order methods.
Possible choices are those from \cref{subsec:secondorder}, \eg \met{SE}{2}{2}, and we obtain
\begin{equation}
 U_V(t^{(CHR4)}_{i_{TV}} \tau) = \prod \limits_{j_V} \prod_{k=1}^{N_V} U_{V_k}(c^{(S2)}_{j_V} t^{(CHR4)}_{i_{TV}} \tau) \prod_{k=N_V}^1 U_{V_k}(d^{(S2)}_{j_V} t^{(CHR4)}_{i_{TV}} \tau).
\end{equation}
Using an HST of order $q=3$ this means we have the leading effects of the non-commutativity among $V_i$
at order three, at order four we find 
the integration error of the HST. At order five we have the non-commutativity between $T$ and $V$ and at order 7 we find the effect
of the non-commutativity among $T_i$.
Of course, the resulting method is only of second order in $\Delta \tau$ but depending 
on the problem it might be preferable to suppress certain contributions to the error.
Instead of sorting the operators into groups it is also possible to use the technique used in \cref{app:hocbm}
directly on the set of operators $\{H_i\}$.

\subsection{Other directions}
\subsubsection{Stability}
We did not really touch the notion of stability so far which turns up in two places in the AFQMC method.
First, there is the question of numerical stability since the algorithm is inherently unstable 
due to the occuring long chains of exponentials in eq \cref{eq:partfuncfinal}.
The way \cite{ALF2017} to mitigate this is to have a regular grid of stabilization points where
intermediate results are compared.
The subdivision of a time step into splitting stages now provides opportunities for additional
stabilization points on a sub time slice granularity.
Second, there is the validity of the approximation to the partition function $Z$ determined by the
linear stability of the splitting method such that all solutions obtained stay bounded \cite{Blanes2008}.
We did not explore that topic here since we were limited by the intrinsic instability
of the AFQMC method which bounded the maximum meaningful $\Delta \tau$ of the second order methods that we could explore.

\subsubsection{Alternatives to the HST}
A lot of the headaches for going to higher order stem from the property 
that the HST induces the semigroup structure. Alternatives to this would therefore 
be of interest.

\subsubsection{Parameter optimization}
The additional degrees of freedom present due to the splitting coefficients
could again be used for the optimization of various properties.
Finding the minimum norm variants of the new complex hermitian methods considered 
in this paper is something that has to be done in future work.
Additionally, the optimization of the average sign comes to mind.
In the realm of second order methods it would be interesting to search for families
of splittings that have positive coefficients but use
the additional degrees of freedom from additional stages for the purpose
of minimizing the discretization error, similar to what has been done for Suzuki's 
fourth order methods in \cite{McLachlan2002}.

\subsubsection{Processors and projectors}
Except for \cref{app:euler} we did not give much attention to the notion of a processor $P$ \cite{Blanes2005, Blanes1999},
since the partition function,
\begin{equation}
 Z^X = \Tr\left( P^{-1} U_H^X(\beta) P \right),
\end{equation}
is invariant under the associated transform by the cyclicity of the trace and hence the Monte Carlo 
sampling will not profit from it.
Nevertheless, it might be worthwhile to explore this structure for obtaining
improved estimates of observables since \cref{eq:errobs} leaves open the magnitude of the 
relative errors in the general case.
A similar idea, that has been studied analytically in \cite{Hansen2009},
is to use a projection $S$ instead of a similarity transform.
A particular case was considered in Ref.~\cite{blanes2010splitting} where complex
symmetric methods were studied for ODEs and it was numerically observed that 
discarding the imaginary part after every time step the 
exponential growth of the error substantially improved.
Since the default updating scheme in ALF is sequential updates,
the notion of an arrow of time is present;
hence this amounts to inserting the projection $S$ onto the real line after every full time step and performing
the truncation at the respective points in the updating process.

\section{Conclusion}
\label{sec:conclusions}
We have applied alternative second order as well as novel higher order splitting methods to the AFQMC method.
We have given evidence 
that the classical Euler splitting is inferior and have provided numerical evidence in \cref{subsec:secondorder}
that there are gains in efficiency to be had when one of the second order methods with more stages is employed:
the methods \met{SE}{2}{2} and \met{S}{2}{2} cost twice as much as one step of \met{S}{1}{2} but 
produce a systematic error that is three times less.
With regard to the existence of higher order AFQMC methods
we have seen that the Goldman-Kaper bound forces us to generically consider complex splittings.
We have numerically shown that the newly introduced family of complex hermitian methods where
one coefficient set is real shows superior behaviour compared to other fourth order methods
with regard to the average sign,
at the price of paying an artificial sign problem. This will limit the potential
use-cases of higher order methods to problems where the pain due to the systematic discretization
error is greater than the pain due to the average sign.

\section*{Acknowledgments}
I thank the DFG for funding through the SFB 1170 “Tocotronics” under the grant number Z03.
The author gratefully acknowledges the Gauss Centre for Supercomputing e.V. (www.gauss-centre.eu) for funding this project by providing computing time on the GCS Supercomputer SuperMUC at Leibniz Supercomputing Centre (www.lrz.de).
The author gratefully acknowledges the compute resources and support
provided by the Universität Würzburg IT Center and the German Research
Foundation (DFG) through grant No. INST 93/878-1 FUGG.
I acknowledge useful discussions with Prof.~Fakher F.~Assaad.

\appendix
\section{Gau\ss-Hermite quadrature with operators}
\label{sec:ghquad}
In this section we give a short derivation of the discrete HST \cite{GUNNARSSON1997530,ROMBOUTS1998271} employed in
equation \eqref{eq:discHST} and how to generalize it.
The basic problem is how to decompose a quadratic operator $B^2$ occuring in an exponential into linear operators.
To that end we will first assume that $B$ has an eigenbasis $|b\rangle $ so that $B | b \rangle = b | b \rangle$
and $\alpha \in \mathbb{R}$ and hence
\begin{equation}
 e^{\alpha B^2} | b \rangle = e^{\alpha b^2} | b\rangle.
\end{equation}
Now we apply a basic gaussian integral identity
\begin{equation}
 e^{\alpha b^2} | b\rangle = \frac{1}{\sqrt{\pi}} \int \limits_{-\infty}^\infty dx e^{-x^2 -2 x \sqrt{\alpha} b} | b \rangle.
\end{equation}
Next we will approximate this one dimensional integral with a Gauss quadrature rule.
Due to the presence of the $e^{-x^2}$ factor in the integrand,
a method from the Gauss-Hermite type is particularly well suited so that we obtain,
\begin{equation}
 \int \limits_{-\infty}^\infty dx e^{-x^2 -2 x \sqrt{\alpha} b} | b \rangle = \sum \limits_{n=1}^N w_n e^{-2 x_n \sqrt{\alpha} b} | b \rangle + R_N
\end{equation}
with $N$ denoting the number of integration points, their location $x_n$, their weight $w_n$ and the remainder term $R_N$.
From the theory of Gauss-Hermite quadrature we know that $x_n$ are real zeroes of Hermite polynomials $H_N(x)$
and 
\begin{equation}
 w_n = \frac{2^{N-1} N! \sqrt{\pi}}{N^2 \left[ H_{N-1}(x_n)\right]^2}
\end{equation}
are positive weights \cite{Gautschi:2004:OPC}.
We will now calculate the remainder $R_N$. To that end we expand $f(x)=e^{-2x \sqrt{\alpha} b}$ in Hermite polynomials,
\begin{equation}
\begin{split}
 f(x) &= \sum_k \frac{a_k}{\sqrt{\pi} 2^k k!} H_k(x) \\
 \text{ with }& \\
 a_k& =\int \limits_{-\infty}^\infty e^{-x^2}H_k(x) f(x) dx = \sqrt{\pi}(-\sqrt{\alpha} b)^k e^{\alpha b^2}.\\
\end{split}
\end{equation}
Calculating the difference $R_N$ between the exact integral $I$ applied to $f$,  $I[f]$ and the quadrature rule $Q_N$ of order $N$,
$Q_N[f]$ we have
\begin{equation}
\begin{split}
 R_N& = |I\left[ e^{-2 x_n \sqrt{\alpha} b} \right] - Q_N \left[ e^{-2 x_n \sqrt{\alpha} b} \right]| \\
 & = \sum \limits_{k=2N}^\infty \frac{a_k}{\sqrt{\pi}2^kk!} Q_N\left[ H_k \right] \\
 & = \sum \limits_{k=2N}^\infty \frac{a_{2k}}{\sqrt{\pi}2^{2k}(2k)!} Q_N\left[ H_{2k} \right] \\
\end{split}
\end{equation}
by exploiting the property that $Q_N$ integrates polynomials of order $2N-1$ exactly and that $Q_N[H_{2n+1}] = 0$.
Using a bound for $Q_N[H_{2k}]$ from \cite{XIANG2012434} we find
\begin{equation}
 R_N \leq 0.816 \sqrt{2}\pi^{\frac{1}{4}} \sum \limits_{k=2N}^\infty \frac{a_{2k}}{2^k \sqrt{2k!}}.
\end{equation}
Inserting $a_{2k}$ and using the Cauchy-Schwartz inequality to treat the series we find
\begin{equation}
 R_N \leq 0.816 \sqrt{2\pi}\pi^{\frac{1}{4}} e^{\alpha b^2}\left[\sum \limits_{k=2N}^\infty \frac{(\alpha b^2)^k}{2^k} \right]^{\frac{1}{2}} 
                                          \left[\sum \limits_{k=2N}^\infty \frac{1}{2k!}\right]^{\frac{1}{2}}.
\end{equation}
To evaluate the first series we have to assume that $|\alpha b^2| < 2$ and we note that the value of the second series
can be bounded from above by $\cosh(1)$, hence
\begin{equation}
     R_N \leq 0.816 \sqrt{2\pi}\pi^{\frac{1}{4}} e^{\alpha b^2} 
     \sqrt{\frac{2}{2-\alpha b^2}} \left(\frac{\alpha}{2} b^2\right)^N
     \sqrt{\cosh(1)}.
\end{equation}
This is consistent with the geometric convergence with respect to the number of integration nodes expected for entire functions \cite{XIANG2012434, LUBINSKY1983338, AlJarrah1983}
and the error behaviour put forward in the physics literature \cite{ALF2017}.

Recasting the equation for the eigenvalues back into operator form we have the required decomposition
\begin{equation}
\label{eq:errHST}
 e^{\alpha B^2} = \sum_{n=1}^N w_n e^{-2 x_n \sqrt{\alpha} B} + \mathcal{O} \left(\left(\alpha B^2\right)^N \right).
\end{equation}

\section{Error analysis of Euler splitting in the case of ALF-1.0}
\label{app:euler}
ALF in its 1.0 implementation uses the Euler-type splitting method \met{NS}{1}{1} = (1,1) and hence
\begin{equation}
 U^{\text{NS}_1 1}_H(\beta) = \prod \limits_{n=1}^{L_\tau} U_T(\Delta \tau) U_V(\Delta \tau)
\end{equation}
which is first order accurate and should at first glance lead to a first order accurate sampling.
To understand why ALF-1.0 nevertheless achieves 
second order behaviour we have to remember that we require only traces of the time evolution operator.
We see that the trace of $U^{\text{NS}_1 1}$ is actually of second order since we can introduce the processor
$P=U_T(\Delta \tau/2)$, exploit the cyclicity of the trace, and obtain
\begin{equation}
\begin{split}
 Z^{\text{NS}_1 1} & = \Tr \left( U^{\text{NS}_1 1}_H(\beta)  P P^{-1} \right)\\
 & = \Tr \left( U_T(-\frac{\Delta \tau}{2}) \prod \limits_{n=1}^{L_\tau} U_T(\Delta \tau) U_V(\Delta \tau) U_T(\frac{\Delta \tau}{2})\right)\\
 & = Z^{\text{S}_1 2}
 \end{split}
\end{equation}
which is exactly the approximate partition function 
using the second order accurate Strang splitting \met{S}{1}{2} $= ((\frac{1}{2},\frac{1}{2}),(1,0))$
and hence gives us a sampling which is second order accurate \cite{casas2008splitting, DeRaedt1983,Fye1986, Hirsch1982}.
Next we need to determine the error behaviour of an observable $O$ and find
\begin{equation}
\begin{split}
 \langle O \rangle^{\text{NS}_1 1} & = \frac{\Tr\left( U^{\text{NS}_1 1}_H(\beta) O \right)  }{Z^{\text{NS}_1 1}}\\
 & = \frac{\Tr\left( U^{\text{S}_1 2}_H(\beta) U_T(-\frac{\Delta \tau}{2}) O U_T(\frac{\Delta \tau}{2})\right)  }{Z^{\text{S}_1 2}}\\
 \end{split}
\label{eq:processedobs}
\end{equation}
showing us that strictly speaking a similarity transform with $U_T$ would be required to obtain a second order 
accurate result of the observable. Using \cite{hall2015lie,Wilcox1967}
\begin{equation}
\begin{split}
 U_T\left(-\frac{\Delta \tau}{2}\right) O U_T\left(\frac{\Delta \tau}{2} \right) & = \exp\left( -\frac{\Delta \tau}{2}T \right) O \exp\left( \frac{\Delta \tau}{2}T \right) \\
 & =O - \frac{\Delta \tau}{2} [T,O] + \frac{\Delta \tau^2}{8} [T, [T,O]] + \mathcal{O}\left( \Delta \tau^3 \right)
 \end{split}
\end{equation}
we obtain using the method dependent error of an observable, eq.~\cref{eq:methoddependenterror},
\begin{equation}
\label{eq:errorcommutator}
\langle O \rangle^{\text{NS}_1 1} = 
\langle O \rangle^{\text{S}_1 2} -\frac{\Delta \tau}{2} \left\langle [T,O] \right\rangle^{\text{S}_1 2} +\frac{\Delta \tau^2}{8} \langle [T,[T,O]] \rangle^{\text{S}_1 2} + 
\mathcal{O}\left( \Delta \tau^3 \right).
\end{equation}
The term linear in $\Delta \tau$,
\begin{equation}
 \left\langle [T,O] \right\rangle^{\text{S}_1 2} = \frac{\Tr\left( U^{\text{S}_1 2}_H(\beta) [T,O] \right)}{Z^{\text{S}_1 2}},
\end{equation}
vanishes by the argument of Fye \cite{Fye1986}:
the trace of the hermitian quantity
$U^{\text{S}_1 2}_H$ times the anti-hermitian commutator (for hermitian $T$ and $V$) vanishes. 
\begin{figure}
\label{fig:eulervsstrang}
 \centering
 \includegraphics[width=0.48\linewidth]{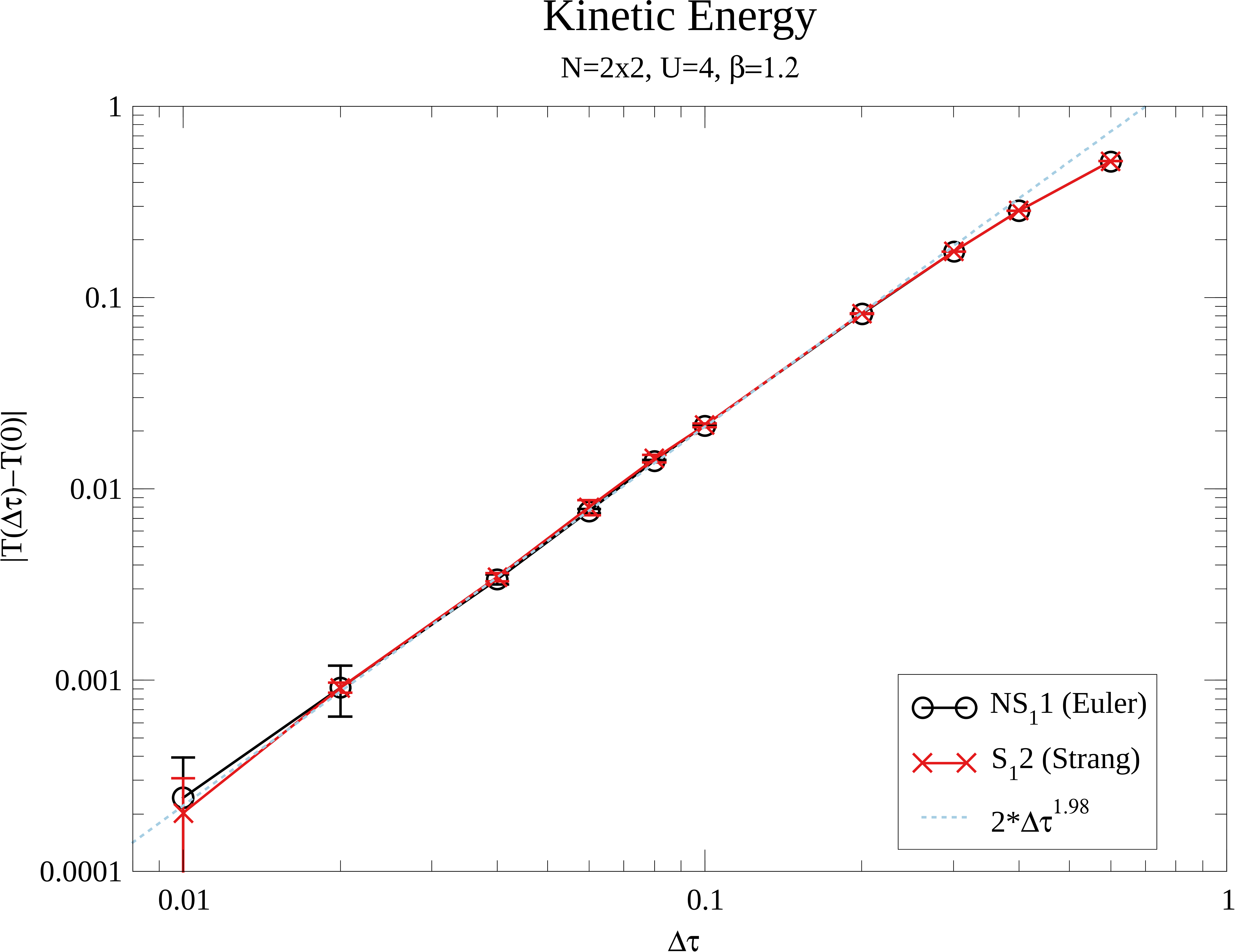}
 \hfill
 \includegraphics[width=0.48\linewidth]{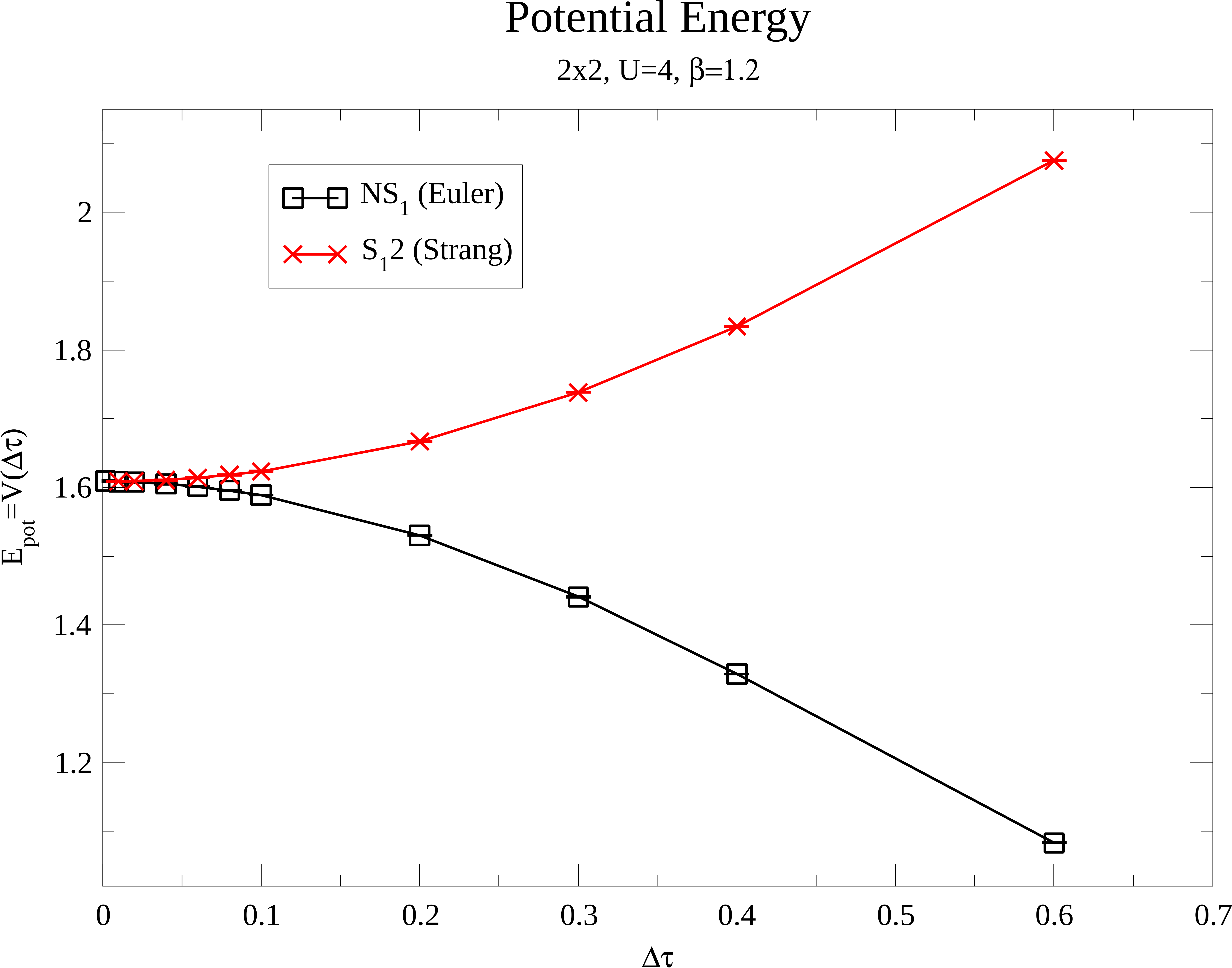}
 \caption{Comparison of the Strang and the Euler splitting with respect to the Kinetic energy $E_{\text{kin}}=\langle T \rangle$ in the left panel
 and the potential energy $E_{\text{pot}}=\langle V\rangle$ in the right panel. We see that the kinetic energy can be reproduced to almost
 identical values for the full range of $\Delta \tau$ and both methods give the expected $\Delta \tau^2$ behaviour.
 On the other hand the right panel shows an increasing discrepancy between the two methods already on a linear scale.}
\end{figure}
Hence the first non-vanishing contribution to the error is 
of second order and gives another contribution to the second order error
of the form $\langle [T,[T,O]]\rangle^{\text{S}_1 2}$.
This additional discrepancy is studied in \cref{fig:eulervsstrang} where we compare the behaviour 
of the Euler method and the Strang splitting on two observables.
First we have in the left panel the kinetic energy $T$ where we see that
the Euler and the Strang splitting show almost indistinguishable behaviour.
In the right panel we compare the potential
energy $E_{\text{pot}}=\langle V\rangle$ and we see the effect of the additional error introduced given in \eqref{eq:errorcommutator}.
We note that this error is present irrespective of the employed Monte Carlo sampling strategy.
This source of error is not present if either the additional similarity transform in \eqref{eq:processedobs}
is done, or the entire Monte Carlo simulation is set up with the Strang splitting, but this is more costly in terms of CPU time.

\section{Higher order checkerboard decompositions}
\label{app:hocbm}
While we did not use the checkerboard method \cite{LOH1992,BaiPropertiesofHubbardMatrices} in the main text since on the one hand we did not want to introduce another approximation and on the other hand 
the test systems would have been too small to have benefitted from the checkerboard method,
we nevertheless introduce here the extension to higher orders in $\Delta \tau$ -
a possibility already mentioned in \cite{Lee2013} -
since it is a very popular, simple and effective approximation to
construct sparse approximations to the exponential of a sparse matrix.
The basic idea is that we split a sparse matrix $H$ into a sum of matrices
\begin{equation}
 H = \sum_i H^{(i)}
 \label{eq:cbsplit}
\end{equation}
in order to be able to approximate its exponential $e^{\Delta t H}$ by applying 
the basic Euler method $E_H(\Delta t)$ to the sum \eqref{eq:cbsplit}
\begin{equation}
 e^{\Delta t H}\approx \prod_i e^{\Delta t H^{(i)}}=:E_H(\Delta t).
 \label{eq:cbeuler}
\end{equation}
The distinctive feature of the checkerboard splitting is that $H^{(i)}$ is constructed in such a way that the sparsity pattern
of $H^{(i)}$ and $e^{\Delta t H^{(i)}}$ is identical, hence
\begin{equation}
 H^{(i)}_{kj} = 0 \Leftrightarrow \left(e^{\Delta t H^{(i)}}\right)_{kj} = 0.
\end{equation}
A method that minimizes the number of used matrices in this case has been given by Lee \cite{Lee2013}.
With the availability of the basic Euler method we can utilise splitting methods to construct higher order approximations
to the matrix exponential. In contrast to the main text we now have full matrix groups and are
hence free to choose arbitrary real splitting coefficients to construct our methods.
At first it might seem that we are limited in the choice of our methods since 
we can only choose methods that can use more than two splitting matrices.
Fortunately, McLachlan \cite{McLachlan1995} has shown that there is a tight connection between the splitting coefficients 
of a method separable into two parts and a composite method that utilizes the 
basic Euler method, \eqref{eq:cbeuler}, of methods with more than two splitting partners that we quickly line out here.
Assume we have a splitting with more than two parts as $E_H(\Delta t)$ in \eqref{eq:cbeuler}.
Then \cite{McLachlan1995, BLANES2002313} has shown that with the coefficients of a method
separable in two parts $X_s p = (\vec{t}, \vec{v})$,
the following method that has the same order can be defined:
\begin{equation}
 \Phi(\Delta t) = \prod \limits_{i=1}^s E_H(c_i \Delta t) E^\dagger_H(d_i \Delta t) 
 \label{eq:symfromtwo}
\end{equation}
where $c_i = t_i - d_{i-1}, d_i=v_i - c_i,$ and $d_0 = 0$. With the help of these equations
every method from the literature given for a splitting in two terms can be adapted to a
splitting given by an arbitrary number of terms.
For the methods considered here we have done this work for the reader and provide tables
collecting all coefficients in the original two-operator representation as well
as the multi-operator representation of \cref{eq:symfromtwo} and the method angle from \cref{def:ma}
at the splitALF project page
\footnote{The precise location is\\ \url{https://git.physik.uni-wuerzburg.de/fgoth/splitALF/-/tree/master/tables}.}
.
In the following we perform two numerical examples
and compare the splitting methods from the main part on a problem that has a full group
structure, and we can evaluate negative time steps.
\subsection{Numerical experiments}
Here we study the quality of the approximation by comparing with the reference value obtained by the full 
matrix exponential $U=e^{\Delta t H}$
and study the dependence of the Frobenius norm of the error matrix $ \lVert \phi(\Delta t) - U(\Delta t) \rVert_F$ with respect to $\Delta t$.\\
\subsubsection{A 1D chain}
\begin{figure}
\label{fig:hex1}
 \centering
 \includegraphics[width=\linewidth]{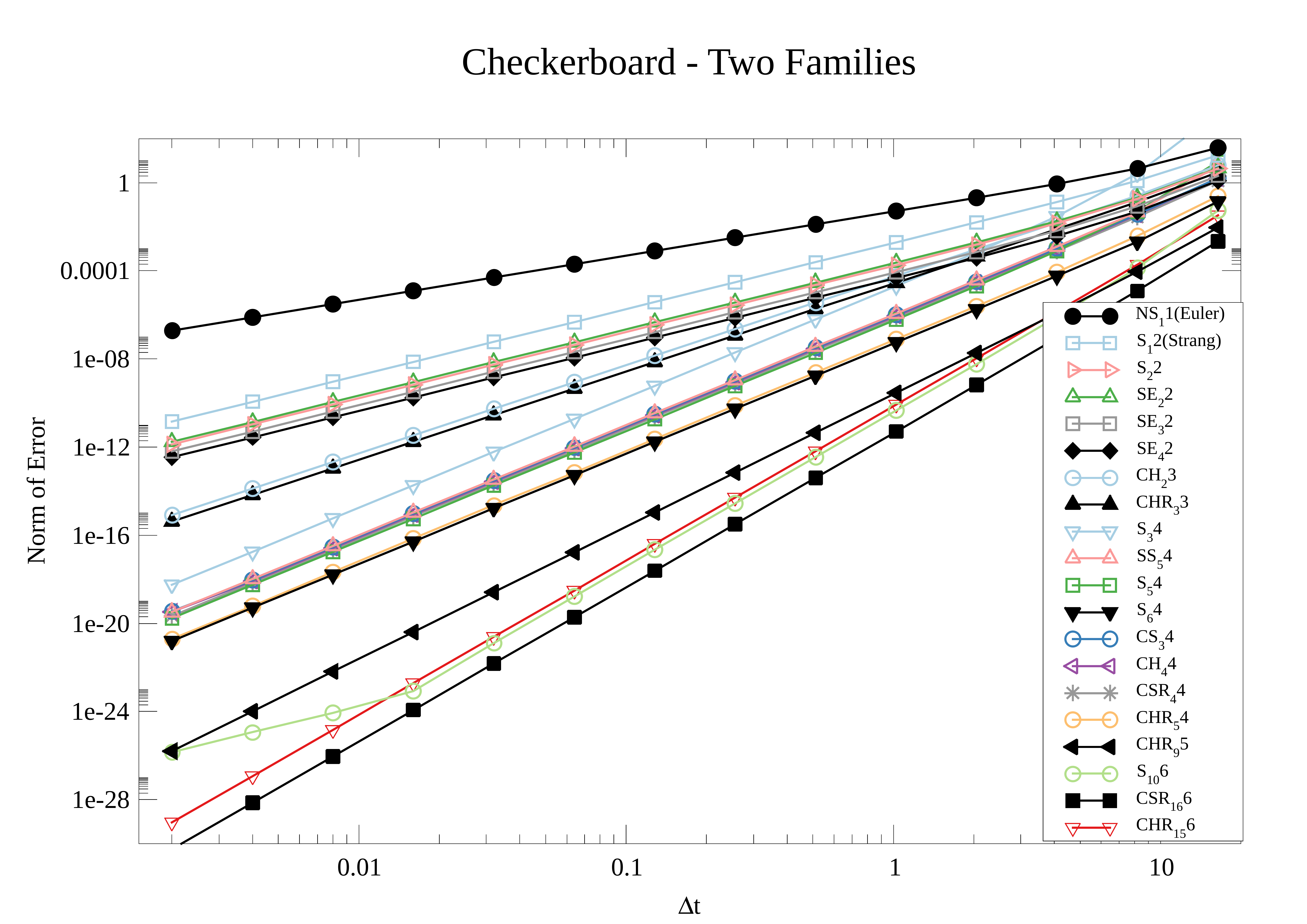}
 \caption{Comparison of the considered methods applied to $H^\text{ex1}$ calculated in 128 digit precision. The norm of the error is the Frobenius norm of the deviation matrix
 $\lVert \phi(\Delta t) - U(\Delta t)\rVert_F$. The kink in the method $S_{10} 6$ is due to insufficient precision in the input coefficients given in the literature.}
\end{figure}

First we consider a small test system describing a one-dimensional chain of length $50$ with nearest
neighbour hopping and open boundary conditions.
$H^{\text{ex1}}$ is given by
\begin{equation}
 H^{\text{ex1}}_{i,i+1} = H^{\text{ex1}}_{i+1,i} = 1. \\
\end{equation}
This matrix can be decomposed into two checkerboard families. The results are shown in \cref{fig:hex1}.
\subsubsection{A randomized matrix}
\begin{figure}
\label{fig:hex2}
 \centering
 \includegraphics[width=\linewidth]{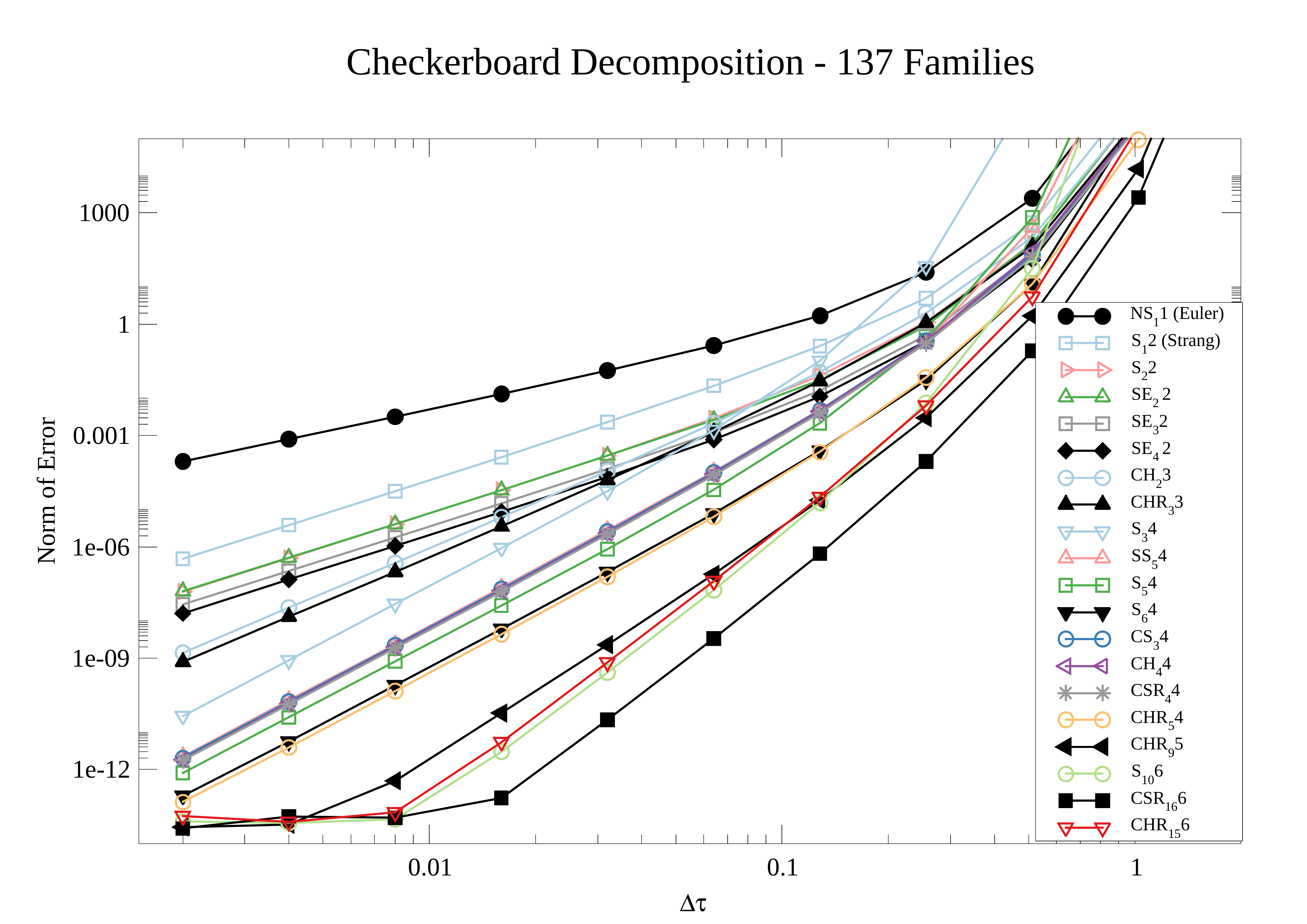}
 \caption{Comparison of the considered methods applied to $H^\text{ex2}$. The flattening out for small $\Delta \tau$
 of the high-order methods is due to the use of the actual machine precision for performance reasons.}
\end{figure}

The previous example was a small system where the basic Euler step only consisted of $2$ applications of an exponential.
Here we consider a different system with $137$ checkerboard families, which results in an equally long Euler step.
The Matrix $H^{\text{ex2}} \in \mathcal{R}^{N \times N}$ with $N=2048$ is constructed according to the following randomized rule:
\begin{equation}
 H^{\text{ex2}}_{ij} = H^{\text{ex2}}_{ji}=
 \begin{cases}
  0.1 +(r-0.95) &\text{if } r > 0.95\\
  0 &\text{else}\\
 \end{cases}
\end{equation}
with $r$ a uniformly distributed random number from $(0,1)$.
In our case this leads to the matrix $H^{\text{ex2}}$ having $210042$
entries and hence $5\%$ of the matrix is occupied. The vertex degree of this matrix is $136$ and the Minimum Split Checkerboard
Decomposition
\footnote{We used our own implementation of the algorithm of \cite{Lee2013}, freely available at\\ \url{https://github.com/CaptainSifff/mscbdecomp}.}
is able to decompose this matrix into $137$ checkerboard families.
The results are shown in 
\cref{fig:hex2}.
\subsubsection{Discussion of the two examples}
We observe that although the considered matrices differ drastically in size and the number of checkerboard families
the results look pretty much the same which is of course due to the fact that matrices are linear operators and results
derived from this property will apply to all matrices.
The family of stabilized second order families \met{SE}{n}{2} by Blanes \etal \cite{BLANES2014} turn out to beat the traditional leapfrog/Strang splitting 
by an order of magnitude and accomplish to be among the most stable even for large values of $\delta \tau$.
The choice of how many stages to use is mostly down to the price of a single Euler step.
Among the fourth order methods the method \met{S}{3}{4} by Yoshida \cite{YOSHIDA1990262,SUZUKI1990319,McLachlan2002} turns out to be 
the worst.
\met{SS}{5}{4} by Suzuki \cite{SUZUKI1990319,McLachlan2002} and \met{S}{5}{4} from \cite{McLachlan1995} behave solid.
The complex fourth order method \met{CS}{3}{4} \cite{Blanes2012} has a slightly smaller error and is slightly more stable for large $\Delta t$.
The second spot among the fourth order methods goes to our \met{CHR}{5}{4} method.
The top performer is \met{S}{6}{4} from \cite{BLANES2002313}, which is almost an order of magnitude more precise
than most other fourth order methods and has stability properties that rival the stabilized second order versions.
In \cite{BLANES2002313} this is attributed
to the fact that for this particular choice of splitting coefficients also higher order contributions have very small coefficients
and hence make it a really solid recommendation if negative time steps are acceptable.
We note that this encouraging result holds although we had to transform a method given for a
two-term splitting via \eqref{eq:symfromtwo}.
The methods of even higher order conform to the expectations: the optimized method \met{S}{10}{6}
displays very small errors, only beaten by \met{CSR}{16}{6}, which is complex and utilizes six stages more.
In that respect \met{CHR}{15}{6} is slightly underwhelming since it utilizes $15$ stages but trails \met{S}{10}{6}.
In this particular example it is hard to find arguments for any particular complex method,
but there could be instances of sparsity where the saving of one stage could outweigh the cost of complex arithmetic as \eg 
in comparing \met{S}{6}{4} and \met{CHR}{5}{4}.

\section{Processing of the order of the HST}
\label{app:suppHST}
In this section we will apply a technique from \cite{BLANES201358} to improve the precision of low order approximations for $V$.
If we employ for the splitting in \cref{eq:split} a second order exact method and now consider an HST with $N=3$ integration nodes,
then by virtue of \cref{eq:errHST} we know that the error of the HST is of 
order $\Delta \tau^3$ and hence would not be usable as an approximation for $U_V(\tau)$.
But since we know the time evolution of the error term,
we can arrange the splitting in such a way that the third order errors in subsequent steps cancel. 
This amounts to imposing the additional constraint
\begin{equation}
 \sum_i v_i^3 = 0
\end{equation}
on the coefficients of a method. Imposing that we were not able to find real coefficients, nor
complex methods with positive $v_i$. But we found a hermitian method that we will consider further.
\begin{figure}
\label{fig:supp_hst}
 \centering
 \includegraphics[width=0.48\linewidth]{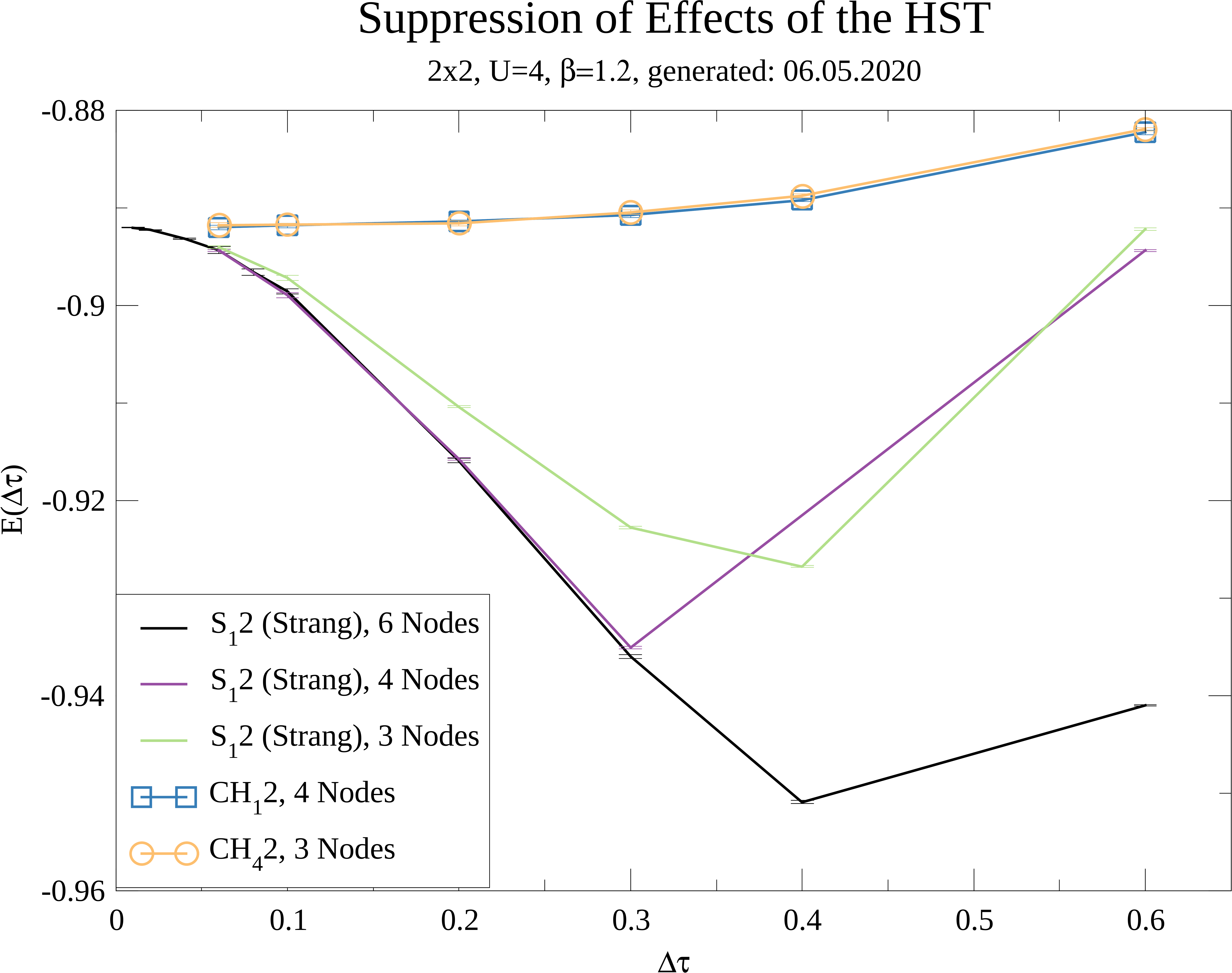}
 \hfill
 \includegraphics[width=0.48\linewidth]{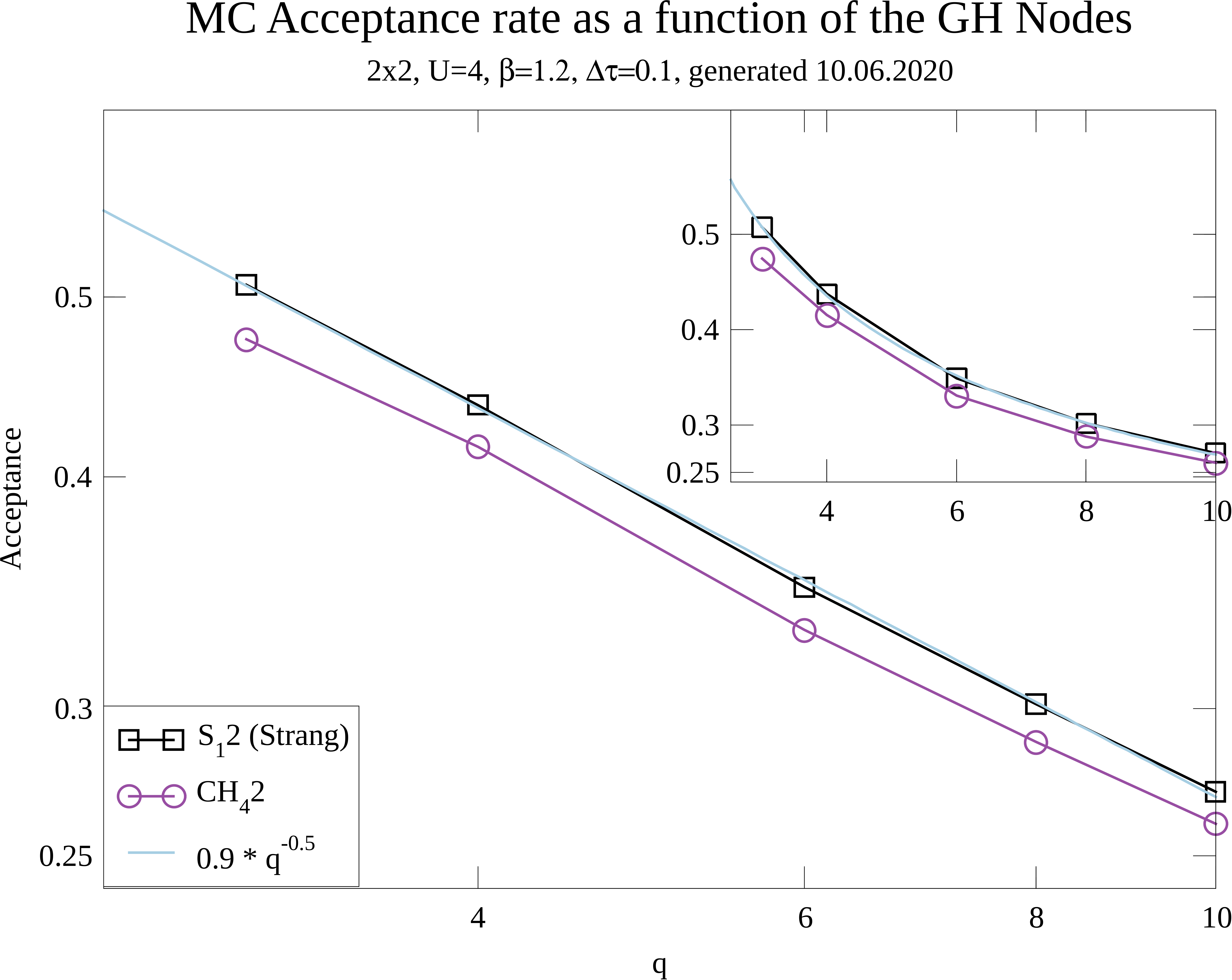}
 \caption{Comparison of the Strang method with the hermitian method \met{CH}{4}{2} that was
 tuned to suppress the second order contributions of an HST of order $q=3$.
 The left panel shows how the new method compares to the Strang splitting. 
 The right panel shows the decay of the Monte Carlo acceptance rate as a function of the order of the HST approximation
 with a double logarithmic scale. The inset shows the same data on a linear scale.}
 \end{figure}
 \Cref{fig:supp_hst} shows how the two methods compare to each other.
 We see that \met{CH}{4}{2} produces
 the same numbers for the energy irrespective of the approximation order $\Delta \tau^q$ of the HST.
 For the Strang splitting we observe that with four and six nodes we get identical numbers whereas 
 the simulations using three nodes deviate visibly.
The right panel gives the Monte Carlo acceptance rate as a function of the number of Gau\ss-Hermite integration nodes
 and the limited data that we have suggests an $N$ dependence of the acceptance $\propto 0.9 N^{-\frac{1}{2}}$.
 A particular nice feature of this approach is that since the nodes are located at $x_{\pm1}=1.2247$ and $x_0=0$ the computational
 cost is comparable to simulations with $q=2$ since the node located at $x_0$ leads to $U_V = \mathds{1}$.
 Sadly, for the considered problem, this method has already an appreciable sign problem
 of $\sigma\approx 0.89 $ that will limit its potential use.
\section{Tables of new hermitian methods with one real set of coefficients}
\label{app:table}
Here we give tables of the new hermtian methods. Note, that the numbers
can also be obtained from the splitALF repository
\footnote{\url{https://git.physik.uni-wuerzburg.de/fgoth/splitALF/-/tree/master/tables}}.
We only give those coefficients which are not determined by hermiticity or consistency.

\begin{tabular*}{0.9\textwidth}{l @{\extracolsep{\fill}} r}
\hline
\hline
\met{CHR}{3}{3}\\
\hline
$t_1$ & $(3-i\sqrt{3})/24$ \\
$t_2$ & $(3+i\sqrt{3})/8$ \\
$v_1$ & $1/3$ \\
\hline
\hline
\met{CHR}{5}{4}\\
\hline
$t_1$ &  $0.07703277090337929685 -i 0.018324799648727332496$ \\
$t_2$ &  $0.20575040027785073837 + i 0.083675123976098550021$\\
$t_3$ &  $0.21721682881876996478 -i 0.152490283366744769770$\\
$v_1$ &  $0.189957023453681608760$\\
$v_2$ & $1/5$ \\
\hline
\hline
\met{CHR}{9}{5}\\
\hline
$\Re(t_1)$ & $ 0.048475520387300861784614942150005$\\ 
$\Im(t_1)$ & $ 0.004320853677325454041666651926455$\\

$\Re(t_2)$ & $ 0.150635519695238479617295632454034$\\
$\Im(t_2)$ & $-0.066664356767339639259626762924050$\\

$\Re(t_3)$ & $ 0.062114931585048261469230020641774$\\
$\Im(t_3)$ & $ 0.128325214677886988240612048515654$\\

$\Re(t_4)$ & $ 0.039425483162945387808087351184557$\\
$\Im(t_4)$ & $-0.151314705015936100913768514487806$\\

$\Re(t_5)$ & $ 0.199348545169467009320772053569630$\\
$\Im(t_5)$ & $ 0.091115040404297316521153563864193$\\
$v_1$      & $ 0.135638579261102446211927963424729$ \\
$v_2$      & $ 0.066861270829291408286939290945765$ \\
$v_3$      & $ 0.119964784754410951531181998568441$ \\
$v_4$      & $ 1/17$ \\
 \hline \hline
 $\text{CHR}^+_{15} 6$\\
 \hline
$\Re(t_1)$ & $ 0.018407829100474781904003178937442$\\
$\Im(t_1)$ & $ 0.060063925504056716987384011897458$\\

$\Re(t_2)$ & $ 0.037640136069562677570517610827614$\\
$\Im(t_2)$ & $-0.078183261287778862724569177453338$\\

$\Re(t_3)$ & $ 0.081602637043566199583289018818188$\\
$\Im(t_3)$ & $ 0.044188227706548767875286813464238$\\

$\Re(t_4)$ & $ 0.081633946450095671181370809036970$\\
$\Im(t_4)$ & $-0.102571562708650305540053523355700$\\

 $\Re(t_5)$ & $ 0.082577837091366965137963780758131$\\
 $\Im(t_5)$ & $ 0.077099467183537374713948506168652$\\
 
 $\Re(t_6)$ & $ 0.096825373607680378408689859741856$\\
 $\Im(t_6)$ & $ 0.035803350489900133270555458990013$\\
 
 $\Re(t_7)$ & $ 0.028447864413379336326027072059858$\\
 $\Im(t_7)$ & $-0.087263167058591631392917553485744$\\
 
 $\Re(t_8)$ & $ 0.072864376223873989888138669819940$\\
 $\Im(t_8)$ & $ 0.147490797656272171753246475176860$\\
$v_1$ & $0.013599999999999999242272785693331$\\
$v_2$ & $0.101321247389239788210399347764220$\\
$v_3$ & $0.084962551210914242400631365647390$\\
$v_4$ & $0.023858373653897122447214292182343$\\
$v_5$ & $0.122034321169012333442504586843360$\\
$v_6$ & $0.094605380895015796981098297399630$\\
$v_7$ & $0.058633436576191442147232738898912$
\end{tabular*}
\bibliographystyle{siamplain}
\bibliography{references}
\end{document}

%% file: splitALFpaper-shared.tex
\usepackage{amsfonts}
\usepackage{graphicx}
\usepackage{epstopdf}
\usepackage{algorithmic}
\ifpdf
  \DeclareGraphicsExtensions{.eps,.pdf,.png,.jpg}
\else
  \DeclareGraphicsExtensions{.eps}
\fi

\DeclareMathOperator{\Tr}{Tr}

\newcommand{\met}[3]{$\text{#1}_{#2}{#3}$}

\newsiamremark{remark}{Remark}
\newsiamremark{hypothesis}{Hypothesis}
\crefname{hypothesis}{Hypothesis}{Hypotheses}
\newsiamthm{claim}{Claim}

\headers{Higher Order Auxiliary Field Quantum Monte Carlo Methods}{F. Goth}

\title{Higher Order Auxiliary Field \\Quantum Monte Carlo Methods
\thanks{
\funding{This work was supported by the DFG through the SFB1170 Tocotronics under the grant number Z03.}}}

\author{Florian Goth\thanks{Universit\"at W\"urzburg, Am Hubland, 97070 W\"urzburg, Germany
  (\email{fgoth@physik.uni-wuerzburg.de}).}
}

\usepackage{amsopn}
